# PACE-SIMS: Checkpoint-Gated Autonomous SIMS Characterization with AI-Agent Quality Control

Anton V. Ievlev[1,*], Heather Hare[2], Yiyang Li[2], Sergei V. Kalinin[3]

[1] Center for Nanophase Materials Sciences, Oak Ridge National Laboratory, Oak Ridge, TN 37831

[2] Materials Science and Engineering, University of Michigan, Ann Arbor, MI 48109

[3] Materials Science and Engineering, University of Tennessee Knoxville, Knoxville, TN 37996

[*] **Corresponding author**

Time-of-flight secondary ion mass spectrometry (ToF-SIMS) is widely used for local chemical investigations across a broad range of materials and systems. However, its operation is expensive in expert time: a trained researcher must supervise the acquisition continuously, tuning parameters as the experiment proceeds, often across a campaign spanning multiple days. Here, we present PACE-SIMS (Pause–Assess–Correct–Execute for SIMS), an agentic workflow that runs a SIMS study as a human–AI collaboration. In this workflow, the researcher specifies the scientific questions and quality requirements, and an AI agent builds the plan and, after approval, executes it autonomously, pausing at checkpoints to judge each measurement and to correct, retry, or escalate. To validate the approach, we applied it to a study of chemical composition in $^{18}O$-enriched $WO_x$ films. During this randomized, blind two-polarity study (8.1 hours, 35 measurements), the agent applied three rule-governed corrections and diagnosed an unanticipated source excursion without requiring event-specific control logic to be programmed in advance. The same run returned transferable measurement science, including a composition calibration and the deposition's tracer-delivery mechanism, from less than two hours of researcher attention. The developed agentic architecture is not specific to SIMS and can be applied to other analytical techniques, with the primary target being destructive measurements, for which optimization-based methods are poorly suited.

# 1. INTRODUCTION

Secondary ion mass spectrometry (SIMS) is among the most information-rich techniques available for local chemical analysis. In its time-of-flight variant (ToF-SIMS), a pulsed primary ion beam sputters the sample surface and the ejected secondary ions are mass-analyzed in parallel across the full spectrum, giving elemental, isotopic and molecular specificity with detection limits that reach the ppm range and, for favorable species, ppb.[1-5] Combined with sputter depth profiling and rastered imaging, this yields three-dimensional chemical maps at nanometer depth resolution and sub-micrometer lateral resolution.[6-8] No other surface-analytical technique combines trace sensitivity, isotopic discrimination and spatially resolved full-spectrum detection in a single measurement. The impact of ToF-SIMS is reflected in publication volume: more than 5,000 publications reporting ToF-SIMS have appeared since 2016, of order 500 per year.[9] Over the past decades it has proven itself across an unusually wide range of systems: from soft matter[10-12] to metals[8,13] and quantum materials,[14,15] from biological specimens[16,17] to microelectronic materials and devices.[18-20]

However, these capabilities are expensive in expert time. A single depth profile usually takes from tens of minutes to hours, while a systematic multi-sample study can occupy an instrument for days. In standard practice a trained researcher supervises the acquisition across the full experiment workflow from selecting species, setting parameters, watching the data arrive, and deciding what to do next. The cost does not end with data acquisition, when reduction, analysis and interpretation of the raw hyperspectral data are themselves time-consuming and associated time requirements frequently exceed the acquisition time.

Unlike many other characterization techniques, ToF-SIMS requires active control of the data acquisition process and does not allow for fixed policy workflows. This stems from two considerations. First, correct acquisition parameters cannot be determined in advance since they depend on sample properties, preponderantly sputter rate, that vary between nominally similar samples (by 50% across the deposition series studied here) and drift within a session. Second, SIMS is destructive, so a measurement acquired with wrong parameters does not merely yield poor data; it sputters away the area of film it measured, so that spot on the sample can never be measured again. Successful experimental orchestration requires rapid human feedback. The value of an experiment is determined not simply by whether the instrument completes the requested measurement, but by whether the resulting data are sufficiently informative to determine the next action. This requires rapid reduction and interpretation of the acquired data, identification of failed or uninformative measurements, and adjustment of acquisition parameters, sampling locations, or even the experimental strategy itself. If this feedback is delayed until post-run analysis, the experiment effectively proceeds open loop: instrument time continues to be spent on measurements whose value has not yet been established. Thus, human expertise must remain closely coupled to the acquisition process, and the rate at which an expert can interpret results and redirect the experiment becomes a fundamental limit on experimental throughput.

The instrument vendors answer the attendance problem with batch automation: scripted measurement queues that execute a predefined list of acquisitions unattended.[21] Batch execution addresses the problem of human presence, but not the problem of experimental judgment. It remains intrinsically open loop: the data are not evaluated between measurements, acquisition parameters are not adapted, and emerging anomalies are not recognized. Consequently, an overnight or weekend queue may be completed successfully from the instrument perspective, only to be found during subsequent analysis to have produced unusable data because a parameter was incorrect, the instrument drifted, or the sample response changed. For destructive measurements, the irreplaceable sample material may be consumed before the failure is recognized. Thus, open-loop automation does not eliminate the need for expert supervision; it exchanges saved attendance time for increased experimental risk.

Over the last five years, autonomous experimentation has advanced rapidly across several fields. Machine-learned policies steer scanning-probe and electron microscopes toward regions of interest, and online analysis closes feedback loops that steer *in situ* characterization at synchrotron beamlines;[22-27] self-driving laboratories close synthesis–characterization loops over robotic platforms;[28-30] and large-language-model agents plan and execute multi-step chemistry campaigns.[31] Most recently, language-model agents have begun to operate instruments directly: a large-language-model (LLM) framework plans, executes and analyzes complete atomic force microscopy experiments;[32] a natural-language companion controls acquisition and analysis at synchrotron beamlines;[33,34] and agentic frameworks are emerging for electron microscopy.[35,36] Closed-loop operation, objective-driven decision-making and natural-language planning are now working laboratory tools.

None of these approaches, however, provides a direct analogue for SIMS or other inherently destructive measurements. Optimization-based autonomous microscopy has largely focused on non-destructive imaging, where the goal is to maximize an objective rather than to establish the validity of every individual acquisition. Self-driving laboratories, in turn, primarily automate synthesis workflows, with characterization serving as one element of a broader loop. LLM-based agents offer much more flexible control, but typically without the structurally controlled, instrument-level authority required when an incorrect action can irreversibly affect the sample or potentially damage the hardware. Similarly existing instrument-operating agents have mostly been demonstrated on non-destructive probes such as AFM, where an unsuccessful measurement can simply be repeated. SIMS does not provide this option. The missing capability is therefore experiment automation for intrinsically irreversible measurement campaigns.

Here we present the development of PACE-SIMS (Pause–Assess–Correct–Execute), a system in which an AI agent and a researcher share control of an unmodified commercial ToF-SIMS instrument across the full lifecycle of a study. The plan is co-created in natural language and validated at two human go/no-go reviews; a state machine then executes it autonomously but checkpoint-gated. At checkpoints, the plan defines the evaluation criteria, while the agent assesses the data and decides whether to proceed, correct parameters, remeasure at a reserved spare location, or escalate to the researcher. Analysis is directed in natural language over the complete persisted raw data, including signals never tracked during acquisition. We should note that much of what the system does could in principle be hardcoded as a fixed policy. However, agentic workflows offer two advantages. In this case, a study is specified in natural language rather than implemented as control logic, so most of the development effort collapses into a planning conversation. At run time, the agent can exercise judgment whereas a fixed rule is brittle and often cannot distinguish a marginal deviation that does not matter from one that does, and recognizing anomalies a script was never written to expect.

We validated the system in a randomized, blind study on $WO_x$ thin films containing a buried $^{18}O$ tracer layer. Three independent quantities were available for these films before the run: film density and thickness from X-ray reflectivity; per-sample sputter rates from an atomic force microscopy crater-depth calibration carried out in a separate operator-controlled session; and the gas-phase isotope ratios fixed by the deposition recipe. Because the films come from a single deposition series with systematically varied stoichiometry, these give a graded reference rather than a single point, so the agent's results could be tested for correct ordering as well as for correct values, and the limits of that reference are known and stated. Tungsten oxide is also a technologically significant material: its oxygen stoichiometry governs resistive switching in memristive devices,[37] electrochemical memory,[38] and electrochromic coloration.[39] Buried isotopic tracers are, in addition, the standard probe of oxygen transport in such films[40], and the depth, thickness and enrichment of a buried layer can be concealed from the analyst without altering the sample. Compositions, mounting order and an independent AFM crater-depth calibration were hidden from the agent until all analysis was complete (*Section 3*). In an 8.1-hour session of 35 measurements in both ion polarities, all four predictions were confirmed against the sealed ground truth. These included exact recovery of the films' enrichment ranking by three independent observables and agreement of the agent-inferred sputter rates with the AFM calibration. Beyond validating the architecture, this study returned transferable scientific results about properties and stoichiometry of the studied $WO_x$ films.

# 2. THE PACE-SIMS SYSTEM

## 2.1 Lifecycle and collaboration modes

The PACE (Pause–Assess–Correct–Execute) architecture organizes a SIMS study as a three-phase lifecycle *(Figure 1)*: Planning, Run, and Analysis & Presentation under control of the two main actors: the researcher and the AI agent. The workflow seamlessly switches the distribution of authority between human and AI. Each phase operates in a distinct, explicit human–AI collaboration mode (co-creation during Planning, bounded autonomy during the Run, directed execution during Analysis & Presentation). The researcher's role shifts correspondingly from co-designer to supervisor to director. The basic unit of operation is the **study**, as opposed to the individual measurement. A study is defined by a set of scientific questions to be answered across one or more samples and replicates, together with explicit requirements on data quality. The researcher specifies these questions and quality criteria in natural language, and the agent translates them into the required measurements, acquisition parameters, and analysis steps. Success is defined by whether the scientific questions have been answered at the required level of confidence and quality, rather than simply by whether a prescribed set of acquisitions has been completed. Importantly, the process is not necessarily linear. Results obtained during the study can generate new questions, modify the sampling strategy, or require additional measurements, feeding back into the planning stage. This creates an outer scientific-method loop around the inner instrument-control loop described in *Section 2.3*. The three phases of this workflow are described below.

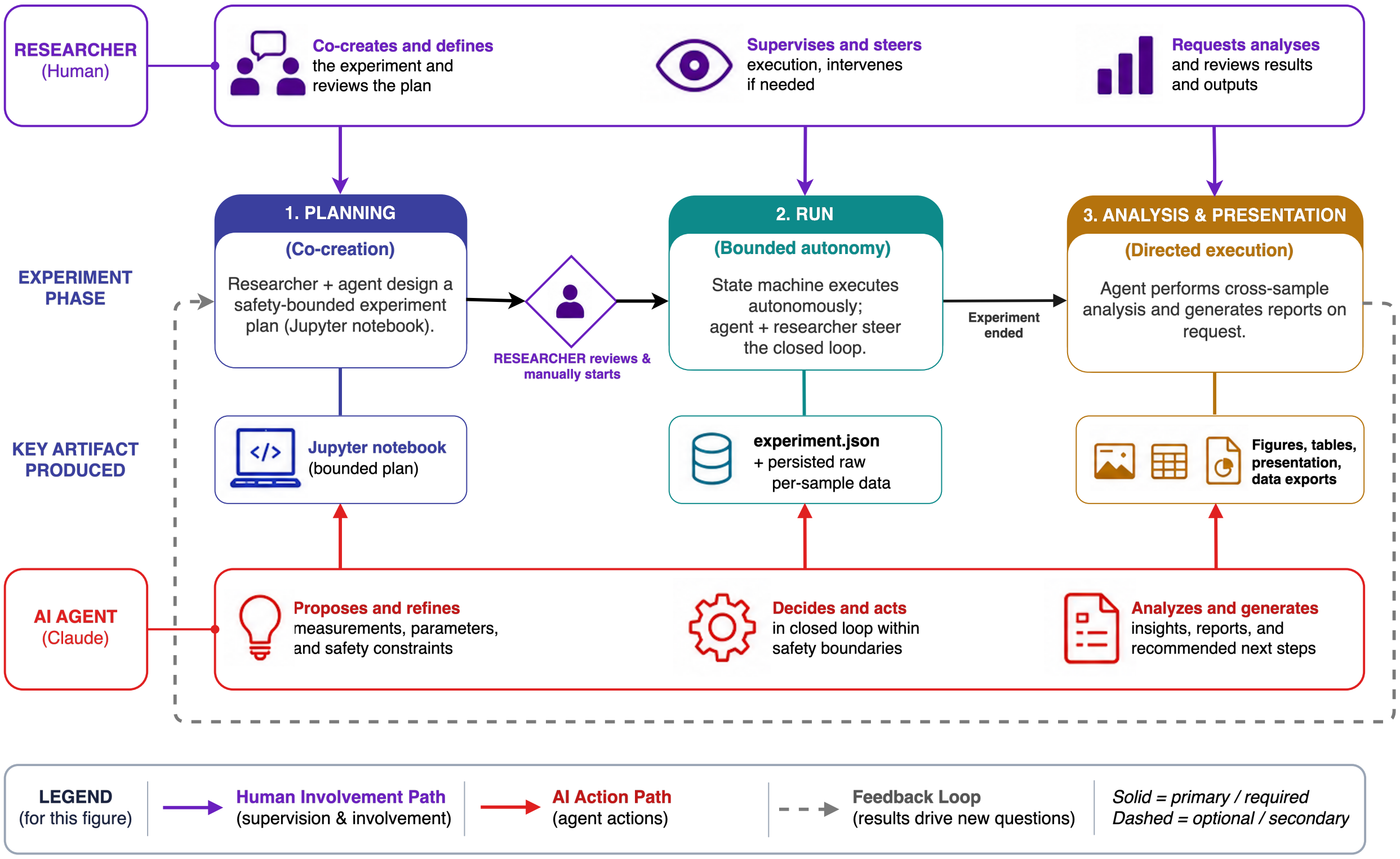

**Figure 1**. The PACE-SIMS experiment lifecycle. A study proceeds through three phases: (1) human-controlled Planning; (2) checkpoint-gated autonomous Run; (3) directed Analysis & Presentation, with the researcher and the AI agent persisting across all three in distinct collaboration modes. Results feed back into planning, closing the outer scientific-method loop around the run-time control loop.

## 2.2 Human-controlled planning

As with any research project, the work starts with planning (*Figure 2*). The researcher describes, in natural language, the scientific question, the samples, the quantitative data-quality requirements, and the corrective-action policy the run should follow. The agent then translates this into a candidate measurement plan (including its own selection of detected species and acquisition parameters, grounded in a curated knowledge base of instrument procedures), and the plan is validated at two human go/no-go reviews operating at distinct abstraction levels. At the first go/no-go review, the researcher examines scientific intent: "are these the right measurements to answer the questions?" The plan is revised through the dialogue until it is approved. After that, the agent renders the plan in executable form: a state-machine definition (in the present implementation, a *Jupyter notebook*) carrying three controls: safety limits, measurement parameters, and the measurement step sequence. At the second go/no-go review, the researcher examines this artifact directly, safety limits first: "is this safe to run on the instrument?" Finally, the researcher manually launches execution and nothing runs autonomously until both reviews have been passed.

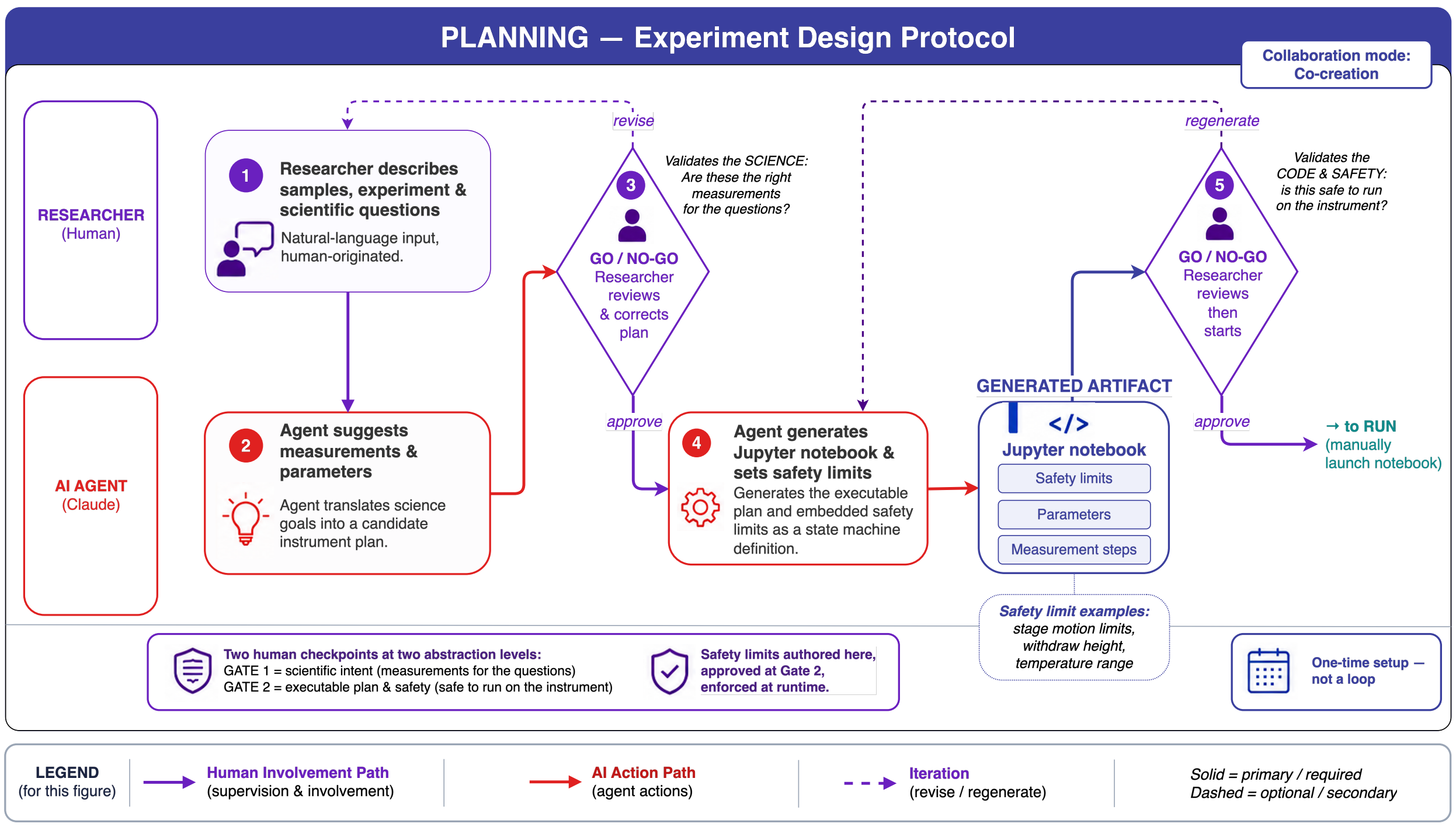

**Figure 2.** Planning phase architecture. Natural-language scientific intent (1) is translated by the agent (2) into a measurement plan (scientific intent review 3), then into an executable state-machine definition (4) carrying safety limits, measurement parameters, and the step sequence (safety and executability review 5). Safety limits follow the provenance chain agent-authored → human-approved → machine-enforced; execution is launched manually only after both go/no-go reviews pass.

This protocol establishes what we term the safety-limit provenance chain: limits are authored by the agent during plan generation, approved by the human, and enforced mechanically by the state machine at runtime (*Section 2.3*). The agent's runtime autonomy is thereby bounded by a contract the human signed before anything ran.

The plan draws its measurement positions from a pre-registered set obtained by optical registration of the sample holder before the run. Alongside the primary measurement sites, the registry reserves sacrificial locations for parameter tuning and spare locations for corrective remeasurement, converting the destructiveness of SIMS, where a failed measurement cannot be repeated in place, into a planned, budgeted resource.

## 2.3 Checkpoint-gated autonomous execution

The Run phase (*Figure 3*) executes the approved plan as a sequence of measurements punctuated by checkpoints. Here, we define a checkpoint as a pause requested in the plan, at which the state machine waits and advances only when an explicit decision is provided by the agent. The placement of these checkpoints is itself part of the experimental plan. The most conservative strategy is to evaluate the data after every measurement, providing the tightest control over sample consumption; this is the approach used in the present study. In other cases, however, measurements should be performed as uninterrupted sequence. For example, in a temperature-driven diffusion experiment, the prescribed temporal spacing between acquisitions carries scientific meaning, and interrupting the sequence for intermediate evaluation would alter the experiment itself. The series should therefore proceed uninterrupted, with the checkpoint applied to the sequence as a whole.

At each checkpoint, the agent retrieves reduced data representations from the analysis service, such as mass spectra, depth profiles, or ion maps, and evaluates them against the quality requirements specified in the approved plan. These requirements fall into two categories. Hard criteria define conditions whose violation automatically invalidates the measurement, whereas advisory criteria identify deviations whose significance requires contextual judgment. The agent therefore exercises discretion only where the experimental plan explicitly permits it. Initial parameter optimization does not require a separate operational mode. It uses the same feedback loop, with convergence criteria replacing the production QC criteria, and can therefore be viewed simply as the first application of the same standing control law.

The decisions made at each checkpoint fall under one of four categories fixed at the planning phase. (i) Accept: release the next measurement unchanged. (ii) Adaptive parameter correction: a quantitative metric outside its acceptance band indicates that the effective acquisition conditions have changed; the agent recomputes the implicated parameter by a predetermined correction law (in the present implementation a proportional scaling, new = current × measured/target, though other laws can be substituted). It then marks the affected measurement invalid, orders a remeasurement at a reserved spare location, and carries the corrected value forward to all subsequent measurements. (iii) Bounded retry: a quality failure with nominal parameters is treated as a transient anomaly and remeasured once at a spare location with unchanged parameters. (iv) Escalate and hold: exhausted retries or an unclassifiable anomaly leave the state machine paused pending human input.

This list bounds the agent's autonomous choices but leaves the researcher's range of decisions open. At any time, the researcher may intervene with natural-language instructions, which the agent translates into the same decision framework: releasing or holding execution, modifying parameters, adding or removing measurement steps, overriding an agent decision, or resolving an escalation. The researcher's runtime authority is limited only by the plan's safety envelope, and category (iv) closes through the human: an escalated checkpoint is resolved by instruction, after which autonomous operation resumes.

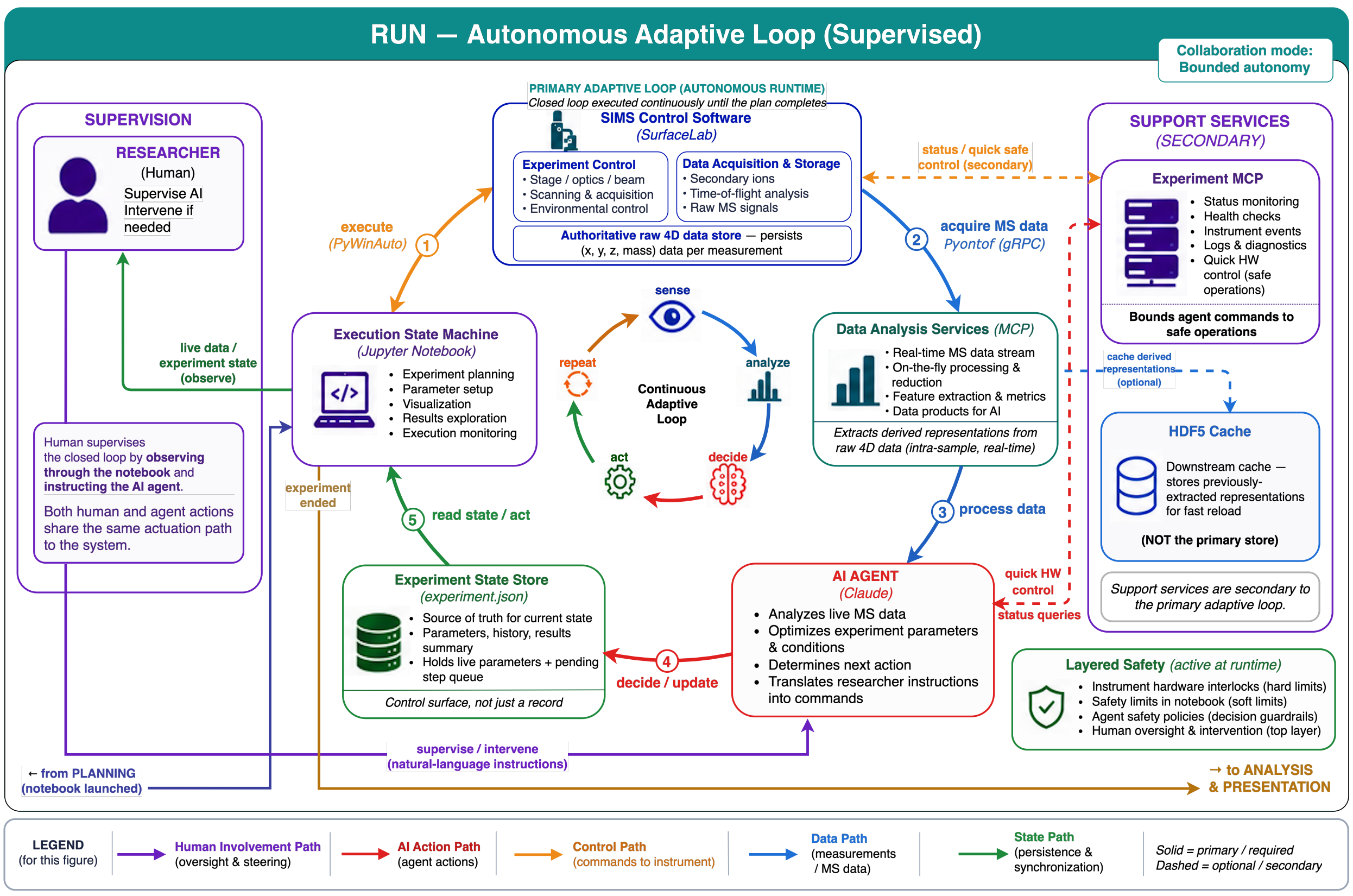

**Figure 3.** Run-phase architecture. The state machine executes the planned measurement sequence, pausing at the checkpoints defined in the plan; at each checkpoint the agent evaluates reduced representations of the measurements just completed against the approved quality criteria and writes exactly one of four decisions (accept, adaptive parameter correction, bounded retry, or escalate-and-hold) to the experiment-state store, the single actuation path through which all human and agent commands reach the instrument. Four independent safety layers bound execution. PyWinAuto and Pyontof (python package for SurfaceLab data access) label the two access paths to the vendor software: PyWinAuto issues control commands through its user interface, while Pyontof retrieves the acquired mass-spectrometry data over a gRPC interface.

All runtime changes, autonomous and human-initiated alike, traverse a single actuation path: agent → experiment-state store → state machine. The researcher never actuates the instrument directly during autonomous operation. The state machine, not the agent, arbitrates when a requested change can be applied safely. One path yields one audit trail and uniform safety enforcement regardless of which actor issued the command.

Execution is de-risked by four independent safety layers: hardware interlocks, the plan's soft limits enforced by the state machine, command bounding at the instrument-services interface, and human oversight. Abort semantics range from a graceful stop, which completes the current step losslessly, to an immediate stop, which forfeits at most the measurement in progress. Note that the authority to stop the run autonomously is granted to the agent per run by the human operator rather than by default.

## 2.4 Data architecture and directed analysis

The decision making during the experiment and post-experiment analysis are determined by how the measurements are stored. Here, we build the architecture where each measurement is retained in two forms (*Figure 4*). The primary record is always the complete instrument-native dataset. For ToF-SIMS, this is a large four-dimensional hyperspectral dataset containing spatial position, sputter cycle, and mass-to-charge information, stored exactly as produced by the vendor software.

The agent does not work with this raw dataset directly. Instead, an analysis service extracts the representations needed for a particular decision, such as an averaged mass spectrum, a depth profile for selected species, or an ion map. Once calculated, these reduced representations are cached, so they can be reused rapidly during checkpoint evaluation or subsequent analysis. Importantly, the cache contains only the analyses that have already been requested. If a new question arises later—for example, a depth profile for a species that was not considered during the experiment—the corresponding representation is simply recalculated from the original raw data. Thus, the experiment is not limited by what the agent happened to examine in real time. The complete information content of every measurement remains available for future analysis.

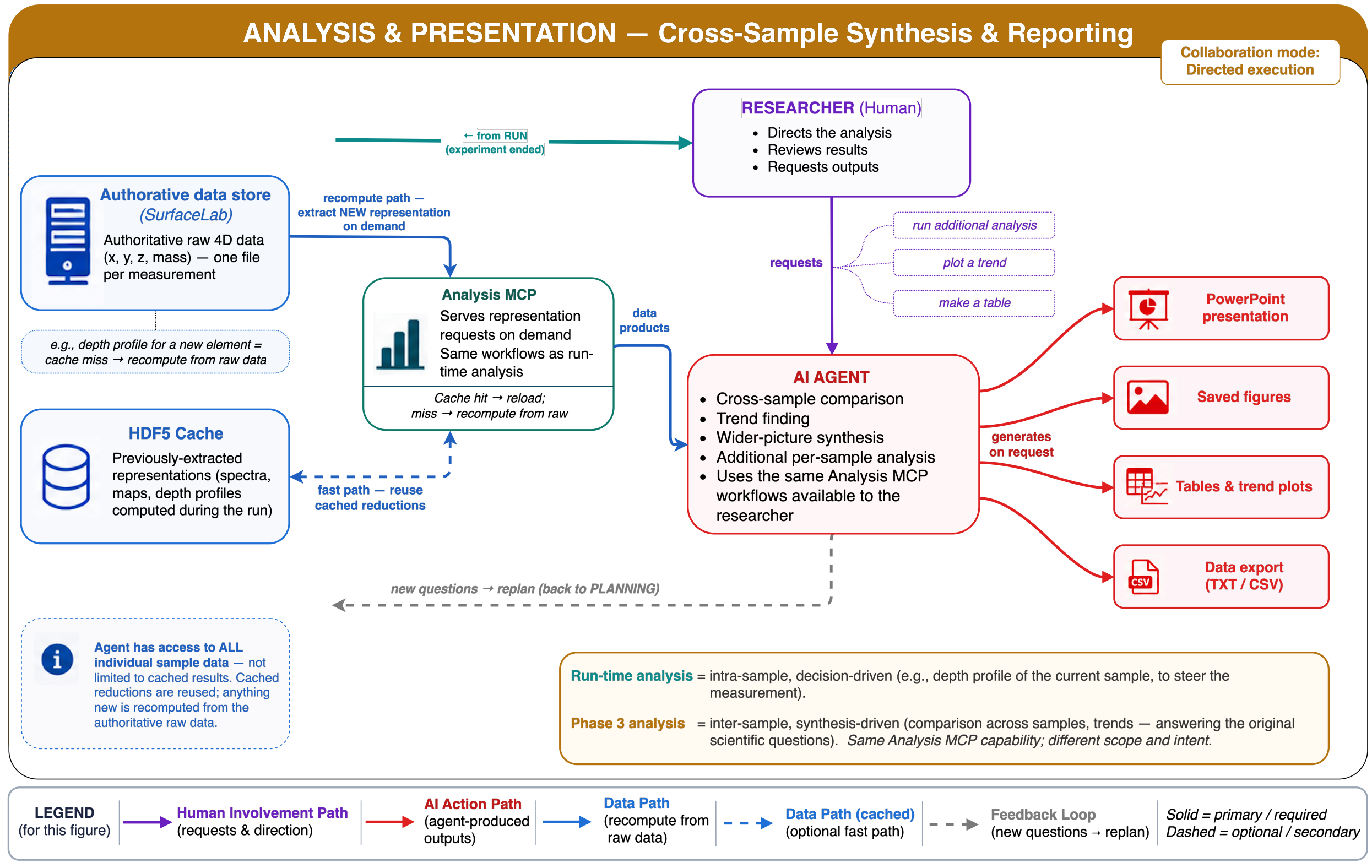

**Figure 4.** Analysis & Presentation phase architecture. The instrument-native store holds the authoritative raw four-dimensional dataset of every measurement; an analysis service extracts reduced representations on request; the cache retains only representations already computed. A request for an un-cached representation, a cache miss, is recomputed from the raw store, enabling re-interrogation of completed runs without additional beam time.

At run time, this division keeps checkpoint evaluation fast: the agent reasons over kilobyte-scale reductions rather than gigabyte-scale raw data. After the run, it is what makes a completed study re-interrogable, as hypotheses formed during analysis can be tested against signals that were never part of the live acquisition, at zero additional beam time.

The Analysis & Presentation phase draws on the same service. The researcher directs in natural language and the agent retrieves or recomputes representations, performs cross-sample synthesis and statistical treatment, generates figures, and assembles reports over the complete persisted study.

## 2.5 Implementation

To implement the architecture described above, we built PACE-SIMS around a commercial ToF-SIMS instrument (TOF.SIMS[5] - NCS, IONTOF GmbH) operated through its vendor software (SurfaceLab 7.5), with no hardware modification: every command the system issues is one an operator could issue by hand. The state machine is a custom Python implementation hosted in a Jupyter notebook, which drives the vendor software through internal COM servers and pywinauto UI automation. The experiment state lives in JSON files that both the state machine and the agent read and write. The agent itself is a large language model (Claude Opus 5), and it reaches the system exclusively through three Model Context Protocol (MCP) services: an experiment service exposing the bounded instrument operations together with the state store, an analysis service performing the data reductions of *Section 2.4*, and a curated knowledge base of instrument procedures and facility methods. Confining the agent to these interfaces is what makes the command bounding of *Section 2.3* enforceable in practice, since capabilities absent from the services cannot be invoked whatever the agent proposes.

# 3. BLIND VALIDATION STUDY DESIGN

In the validation study, we used four tungsten oxide ($WO_x$) thin films from a single reactive-sputter deposition series. Each film used a different reactive gas composition, with the total gas flow fixed at 40 sccm (standard cubic centimeters per minute, equivalent to $6.7 \times 10^{-7}$ m$^3$/s; sum of $O_2$ and Ar flows): 2 sccm of $O_2$ (5%), 4 sccm (10%), 6 sccm (15%), and 8 sccm (20%). The low-oxygen films are substantially oxygen deficient while the high-oxygen films are essentially fully oxidized $WO_3$. The film densities, measured independently using X-ray reflectivity (Supplementary Information, *Figure S1*), ranged correspondingly from 10.30 to 6.51 g/cm$^3$. Each film is nominally 45 nm thick on a silicon substrate, capped with approximately 30 nm of $SiN_x$ by plasma-enhanced chemical vapor deposition. Within the 45 nm $WO_x$ film, a buried ~15 nm layer near mid-film is $^{18}O$-enriched, produced by replacing 1 sccm of the natural-abundance $^{16}O_2$ flow with 99% isotopically enriched $^{18}O_2$. Growth and metrology details are given in the Methods section.

Blinding here serves to test the system rather than to establish the material results, which rest on conventional measurements: a randomized blind study with predictions stated in advance and tested against sealed ground truth is the strictest available demonstration that the agent's decisions followed from the data in front of it rather than from expectations supplied to it. The study was blind from the agent's side: neither the compositions nor their assignment to holder positions was available to it. The information provided to the agent was that the holder carried four samples of the same nominal stack, drawn from one deposition series with differing deposition conditions, each containing a buried $^{18}O$-enriched layer of unknown depth, thickness and enrichment. On that basis it was asked three questions: the depth, thickness and $^{18}O$ fraction of the enriched layer at each position; a ranking of the positions by enrichment together with any other systematic differences it could detect; and how the yields of the characteristic species depend on the sample in each of the two ion detection modes — the instrument can collect either negatively or positively charged secondary ions, and the two respond differently to chemistry — and hence which mode discriminates composition best. The mounting order was randomized immediately before mounting, so holder position carries no information about composition. Two sets of ground truth existed before the run and were withheld from the agent: the composition assignment (which oxygen flow, and hence which density, sat at which position) and an atomic force microscopy crater-depth calibration of the same four films, acquired in an earlier operator-run session, which fixes the sputtered depth per unit ion dose independently of any SIMS measurement. Neither was used by the agent during the run or during the analysis of its data, and both were compared against the blind-side results only after the run and analysis phases were completed.

The blinding was one-sided by design: the operator knew the mounting order, performed the reference calibration with that knowledge, and stated the predictions afterwards, since it is the agent and not the analyst that the study sets out to test. The agent worked without access to any ground truth, and the blind run sampled fresh locations on each film, away from the calibration craters. The blinding was enforced structurally as well as by instruction, since an instruction not to look is only as good as the compliance of the agent. The run and its analysis were carried out in a fresh agent session with no access to earlier conversations, the curated knowledge base was audited to remove any document derived from previous measurements on these films, and the data services were scoped so that earlier datasets on the same samples were unreachable. The agent was additionally instructed to consult no prior results and to report immediately if any tool returned them. The quality requirements supplied with the study were quantitative: 50 to 60 data points across the $WO_x$ layer, count rates below the ceiling above which the detector no longer responds linearly, 5 counts per pixel per shot, on any channel used for ratios, termination on the signal-triggered stop that detects the substrate rather than on the fixed scan limit that serves as a backstop, and flat interfaces in the depth-resolved maps, with composition differences between samples declared expected and never a failure. The corrective-action policy bound each of these to a single response: a sampling-density or stop-condition violation was corrected by the proportional law, invalidated and remeasured at a spare location with the corrected value carried forward; a transient artifact at nominal parameters earned one unchanged remeasurement; a uniform change in absolute yield with ratios intact was flagged and passed; and saturation, an empty channel, or any second failure at the same location escalated to the operator and held the run.

Four quantitative predictions were stated in advance of unblinding and evaluated against ground truth the analysis could not see; they are listed in the Supplementary Information (*Section S2*). They were that (i) the relative sputter rates inferred by the agent during the run would agree with the sealed AFM calibration; (ii) the mass removed per unit ion dose would be constant across the four films, at the level of constancy already observed on the calibration dataset; (iii) no recalibration would be required in the second polarity block, because the corrections established in the first block transfer to it; and (iv) a drift sentinel acquired at the end of the session would reproduce the measurand within the run's own quality-control tolerance. The four are deliberately different in kind: (i) and (ii) ask whether the agent's adaptive corrections are physically correct rather than merely self-consistent, testing them against independent metrology and against mass conservation; (iii) tests the physical assumption underlying those corrections; and (iv) bounds how much of any difference between samples could be an artifact of when it was measured.

The plan called for 24 main measurements (four samples, two ion polarities, three replicates each) with a target of 50 to 60 data points across the $WO_x$ layer as the sampling criterion the adaptive correction acts on. The two polarities were acquired as separate blocks with a single switch between them, the sputter parameters converged in the first block being carried unchanged into the second, which is the basis of prediction (iii); the session closed with the drift sentinel of prediction (iv). The sequencing and its rationale are described in the Methods.

# 4. RESULTS

## 4.1 Autonomous execution

The run produced 35 measurements over an 8.1-hour session: 24 planned sample acquisitions, six auxiliaries (tuning, quality-control references, mid-block anchors and the closing drift sentinel), and five corrective remeasurements (*Figure 5a*). Thirty of the 35 passed their checkpoints; the five that did not were invalidated by the agent and remeasured at reserved spare locations. Cycle times averaged 11 minutes per measurement in the negative block and 17 in the positive, whose two-phase acquisition is slower. Every dynamically stopped profile terminated on its substrate trigger and the 400-scan backstop was never reached.

Tuning began from a reconnaissance profile and converged on a sputter setting for the first sample within a few iterations, each step applying the proportional correction law of *Section 2.3*. The plan then assumed (as a human operator typically would before an experiment) that one setting would serve all four films. However, this was not the case. The first replicate on the second sample returned 48 points across the $WO_x$ layer against an acceptance band of 50 to 60, and the agent reduced the sputter frames (sputter time per scan) accordingly; the third sample returned 66 points and the frames were increased; the fourth returned 45 and they were reduced again (*Figure 5b*). The converged settings differed by 44% across the four films. Had the single tuned value been carried through unchanged, three of the four samples would have fallen outside the sampling criterion, and the study would have produced non-comparable depth resolution across the very samples it set out to compare.

None of these three corrections was scripted. Each followed from a quantitative criterion evaluated at a checkpoint, and each was applied in the same way, the adaptive-correction path of *Figure 3*: invalidate the affected replicate, remeasure at a reserved spare location, and carry the corrected value forward. One correction was upward, which is worth noting because a controller that only ever reduces an over-shooting parameter would not be exercising a control law but clipping one.

The run also encountered an unexpected instrument fault. Partway through the first sample emission of the $Bi_3^+$ liquid-metal ion gun dropped out mid-acquisition; the agent invalidated the measurement and remeasured at a spare location, as the policy prescribes for a transient anomaly. The repeat returned intensities about 30% low and, recognizing that the deviation was sustained rather than transient, the agent escalated and held, the fourth decision of the run-phase menu (*Figure 3*), rather than consuming further locations. An operator realigned the source, after which the third attempt passed quality control and autonomous operation resumed. The escalation path was therefore exercised in production rather than in rehearsal, and it was reached by diagnosis rather than by exhausting a retry counter.

Two further features of the session should be noted. The mid-block anchor measurements on the first sample and the reversed sample order of the positive block were the agent's own additions during planning, not part of the researcher's requirements. Also, the second polarity block required no recalibration at all, the settings converged in the first block carrying through unchanged. Attended operator time totaled 40 minutes of the 8.1-hour session, confined to the initial launch, observation of the first few measurements, and the source realignment. The realignment itself cost about two minutes of instrument time beyond a routine cycle (operation resumed 13 minutes after the escalated measurement completed), and the final 6.7 hours of the session, 28 measurements, ran with no operator intervention at the instrument.

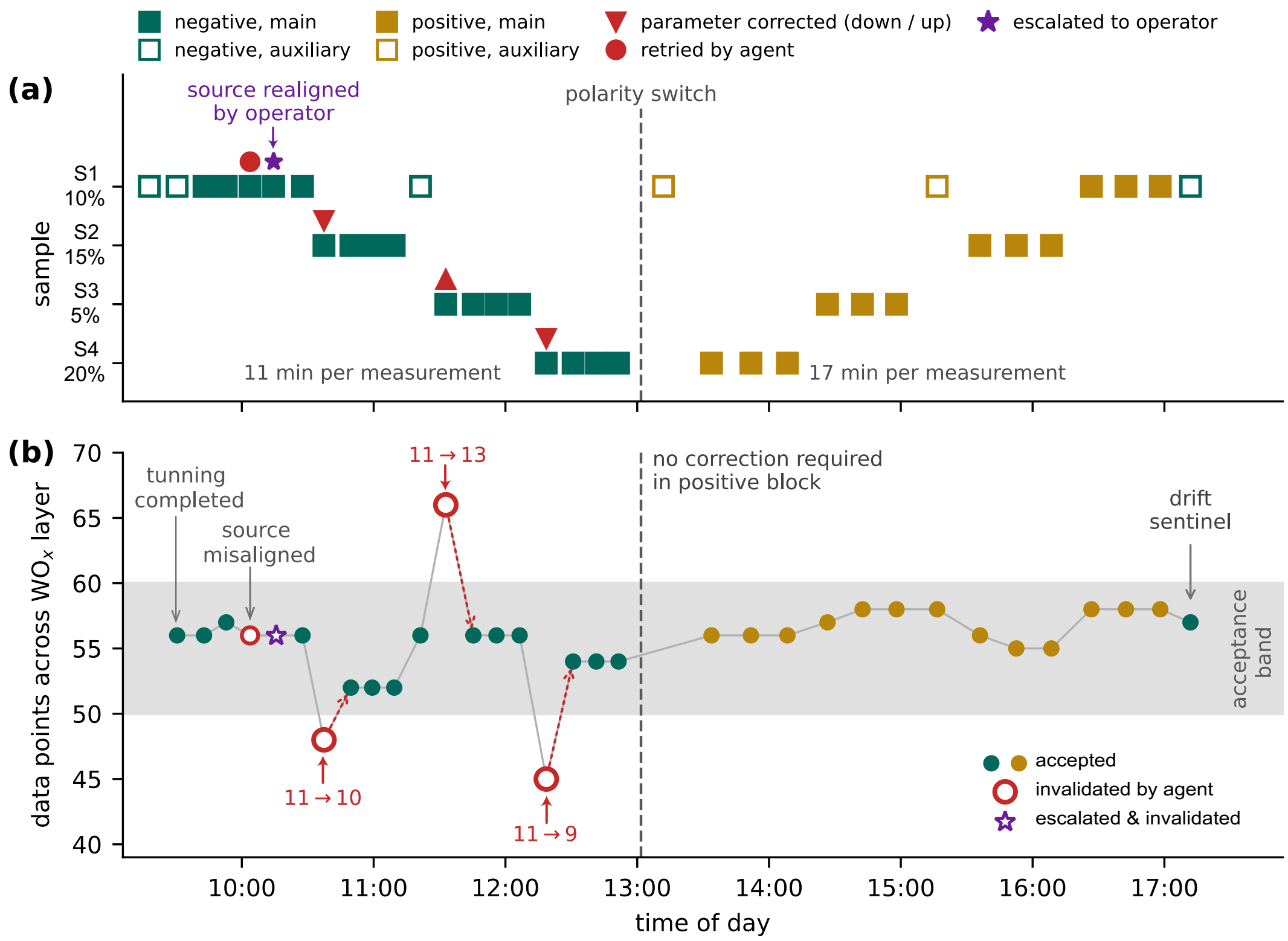

**Figure 5.** Autonomous execution of the blind validation run. (a) Acquisition timeline: one square per measurement, by sample row (labeled with growth $O_2$ fraction) and completion time. Filled squares are main-sequence acquisitions, hollow are auxiliary (tuning, references, drift sentinel); color denotes polarity (teal negative, gold positive). Symbols above squares mark agent decisions: triangles, parameter correction (with sign); circle, retried; star, escalated to the operator. (b) Data points across the $WO_x$ layer per profile against the approved acceptance band (shaded). Open circles are agent-invalidated measurements; the three out-of-band excursions are annotated with the sputter time (frames) correction each triggered, dotted arrows leading to the in-band remeasurement. The corrected settings carried into the positive block, where no correction was required (prediction iii); the drift sentinel closed the session (prediction iv).

## 4.2 Blind-side results

Figure 6 summarizes the experimental results as a function of the sample labels, which were the only sample identifiers available during the blind experiment. The calibrated $^{18}O$ fraction profiles in the two detection modes (*Figure 6a,b*) show a single buried enriched band in every film, the three replicates per sample nearly coincident. The layer sits at mid-film in all four samples, centered at 50 ± 0.7% of the film thickness with an apparent width of 14 to 15 nm: the deposition conditions changed how much $^{18}O$ was incorporated, not where the layer formed or how thick it grew — a useful negative result, since it isolates incorporation as the single variable across the series.

Enrichment fractions over a fixed 10 nm analysis window centered on the layer span a factor of four across the set: 45.7%, 23.1%, 16.0% and 11.4% in the negative ion detection mode, with replicate scatter in the last digit (*Figure 6c*). The positive ion detection mode determinations of the same quantity run systematically higher, by roughly 5%; the offset is visible in *Figure 6c* and is discussed with the two modes below. The volumetric $WO_x$ sputter rate spans a factor of 1.5 across the four films (*Figure 6d*). The SiN cap, nominally the same material on every sample, sputters at a constant rate (within 1.1%) over the same measurements, so the spread in $WO_x$ sputter rate is a property of the films rather than of the beam or the session.

Absolute composition is read from a cluster-ion ratio rather than from the elemental signals, because cluster formation is much less sensitive to the chemical environment of the atom emitted. That ratio, oxygen to tungsten as their cesium clusters ($(^{16}OCs_2^+ + {}^{18}OCs_2^+)/WCs^+$), the positive mode composition observable, spans 6.3 to 24.0 across the set (*Figure 6e*). None of these quantities is ordered in the sample label, so *Figure 6's* x-axis carries no information, by construction. Ordering the samples by any one of them, however, produces the same sequence every time: S3, S1, S2, S4. Whether that shared ordering means anything is what the sealed envelopes decide.

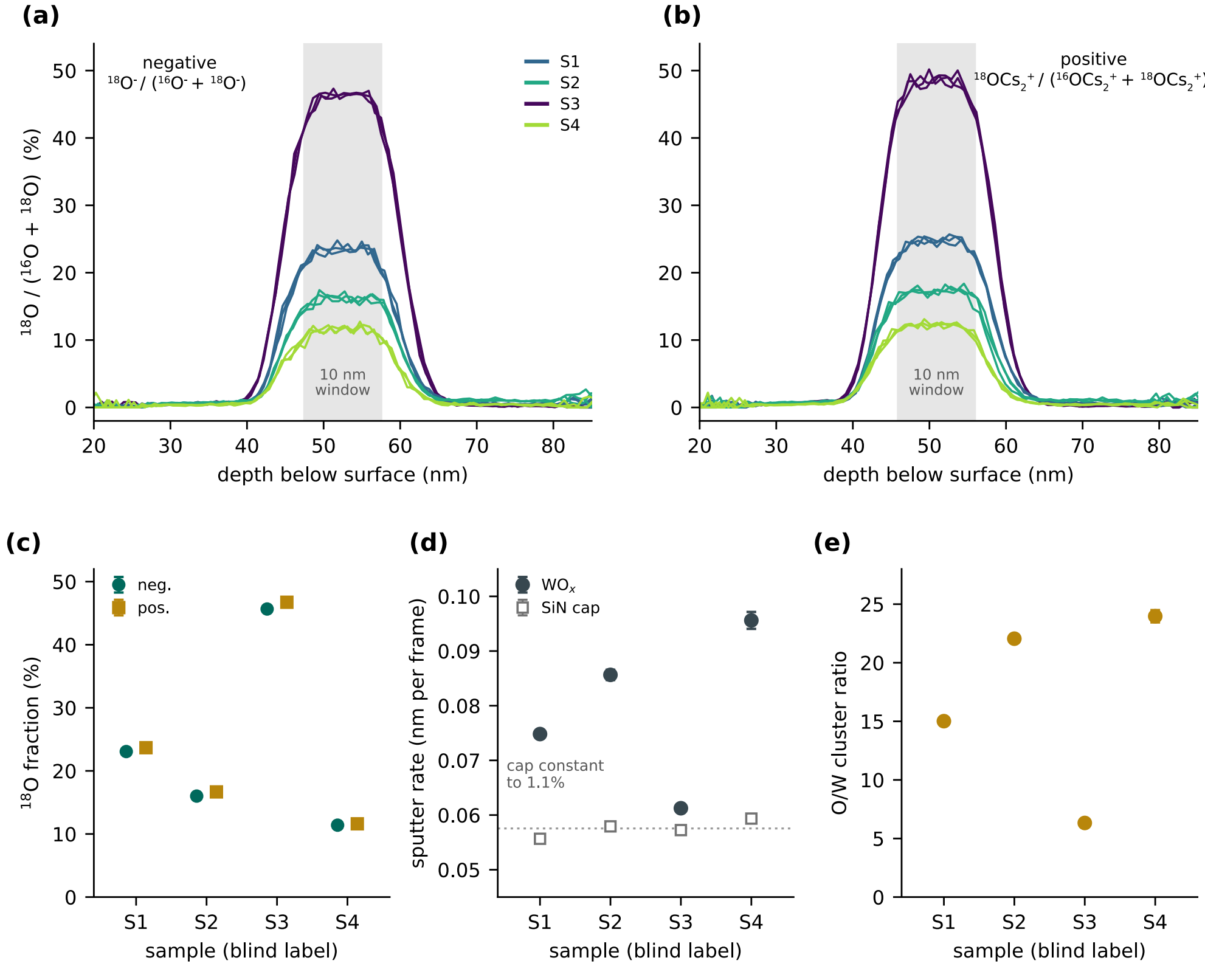

**Figure 6.** Blind-side results, plotted against sample labels, the only identity available to the agent. (a,b) Calibrated $^{18}O$ fraction depth profiles in the negative ($O^-$) and positive ($OCs_2^+$) ion detection modes, three replicates per sample; the shaded band is the fixed 10 nm evaluation window. (c) Windowed $^{18}O$ fraction per sample in both modes; the positive mode reads systematically higher, by roughly 5%. (d) Volumetric $WO_x$ sputter rate per sample; the SiN cap rate (open symbols) is constant to 1.1%, an internal check that the spread is a film property. (e) Cs-cluster O/W composition ratio. Error bars are 1 SD over replicates. No panel is ordered in the sample label; the x-axis carries no information, by construction.

# 5. ANALYSIS AND DISCUSSION

## 5.1 Validation against sealed ground truth

With the blind-side analysis frozen, the sealed composition assignment and the reference calibration were opened and compared against it. The enrichment ranking recovered from the data was exact. Plotted against the revealed $O_2$ flow (*Figure 7a*), all three observables (the $^{18}O$ fraction, the volumetric sputter rate and the cluster composition ratio) are strictly monotonic: the zigzags of *Figure 6* resolve into the deposition series. The 2-sccm $O_2$ film was the most enriched because the reactive sputter gas was 1 sccm of $^{16}O_2$ and 1 sccm $^{18}O_2$, while the 8-sccm $O_2$ film was the least (7 sccm of $^{16}O_2$ and 1 sccm of $^{18}O_2$), with the between-sample separation exceeding within-sample scatter by one to two orders of magnitude. The ordering also reproduces across the two polarity blocks, which visited the samples in opposite order, so it cannot be an artifact of drift during the session.

The agent-inferred sputter rates agreed with the sealed AFM calibration data to 1.05% (*Figure 7b*). This is the comparison that tests the adaptive mechanism rather than its outcome: the agent had inferred relative rates from its own points-per-layer arithmetic during the run, blind, while the calibration had been measured independently by crater-depth metrology in an earlier session (Supplementary Information, *Figure S2*).

The mass removed per unit ion dose, the product of volumetric sputter rate and film density, was constant across the samples within 3.1% (*Figure 7c*), replicating on blind data the level of constancy previously observed on the calibration set. The 1.5-fold spread in volumetric sputter rate across the series is therefore quantitatively a density effect, and the parameter the agent adapted was tracking a physical property of each film rather than an instrumental artifact.

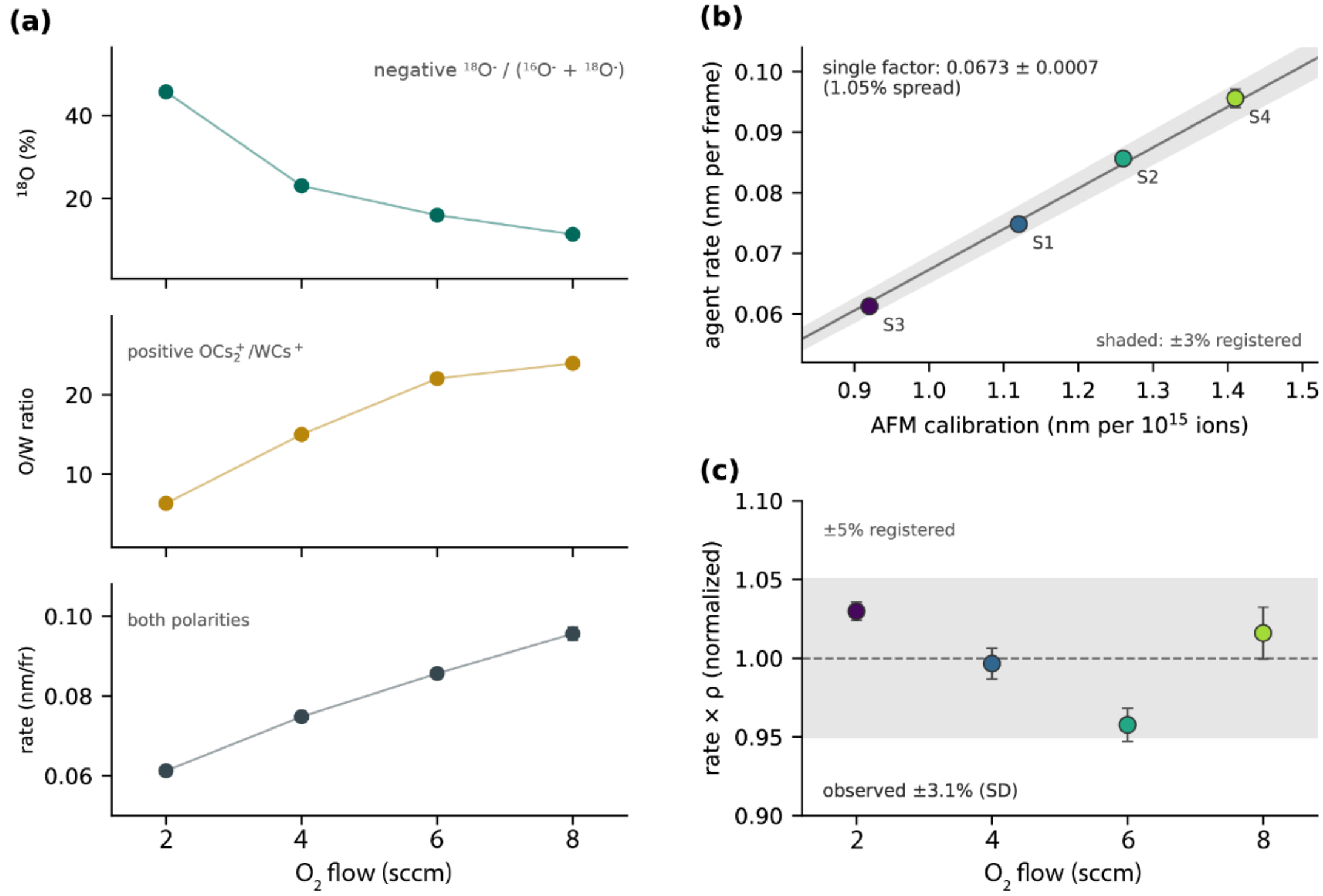

**Figure 7.** Validation against sealed ground truth; the x-axes now carry the revealed deposition parameters. (a) The three observables of *Figure 6* versus $O_2$ flow, each strictly monotonic, recovering the sealed sample order three times over. Error bars are 1 SD over replicates. (b) Prediction (i): agent-inferred sputter rates versus the sealed AFM crater-depth calibration; a single conversion factor fits all four samples with a 1.05% spread, inside the registered ±3% acceptance (shaded). (c) Prediction (ii): mass removed per unit ion dose (volumetric rate × film density, normalized) constant to ±3.1% (SD) against the registered ±5% band (shaded).

The drift sentinel closed the session with the measurand reproducing within 2.5% of the morning reference, while absolute yields over the same interval moved by 12%; across the whole study the first sample's isotope fraction held to 7.8 ± 0.1% through a 32% swing in absolute signal. All four predictions were therefore confirmed against the sealed ground truth, each by a margin well inside its threshold. Two of the films had been measured in earlier operator-run sessions, and the blind values fall within those historical ranges.

The acquired data additionally supported an absolute-composition analysis (Supplementary Information, Section S5). The independently measured film densities served as composition standards against which the Cs-cluster oxygen-to-tungsten ratio was calibrated for all samples. The films span compositions from $x$ = 1.12 to 2.96, and over the oxidized part of that range the ratio is linear in composition (O/W = 8.16 x, $R^2$ = 0.9995); the most metallic film falls about 30% below the trend, attributable to a residual matrix effect — the dependence of ion yield on the chemical environment of the emitting atom. That this analysis was performed entirely after the run, with no additional beam time, illustrates a central benefit of the architecture: the persisted raw data remain open to interrogation long after acquisition is complete.

## 5.2 Adoption and practical considerations

A similar approach can be adopted for autonomous studies with other analytical techniques, and nothing in the architecture is specific to SIMS. What has to be built follows the run phase of the workflow, and only the first part of it depends on the instrument.

The instrument layer is driven by the state machine, and it is the one component that must be designed specifically for each technique. It has to express a set of operations (set the acquisition parameters, start a measurement, report the run state, *etc.*) through whatever control path the instrument offers, whether a vendor scripting interface, an instrument API or automation of the acquisition software. In our experience this is the bulk of the work, and the safety limits belong here as well, stated as explicit guardrails on motion, scan counts and permitted operations and enforced by the state machine rather than by the AI agent.

The agent reaches the experiment only through service interfaces, which supply the reduced representations it judges measurements on and the instrument state it reasons about. What these services expose defines what the agent can see and do, so the reductions appropriate to the technique are chosen here: for SIMS they are spectra, profiles and maps, while for other methods they could be something entirely different. This also bounds the agent, since an operation absent from these services cannot be invoked whatever the agent proposes.

Between the agent and the state machine there has to be a written channel. Here it is a JSON experiment state that both read and write, so that every decision, parameter change and pending step is recorded in one place and the state machine, not the agent, decides when a requested change can be done. So, the agent acts by writing to it rather than by driving the instrument, which makes the run auditable afterwards.

The remaining inputs do not depend on the instrument. The study specification — the scientific questions, the quantitative criteria that define an acceptable measurement, and the action to take when each criterion fails — is written in natural language rather than in code; the one used here is reproduced in the Supplementary Information, and its structure serves as a template. Alongside it sits a set of knowledge entries describing local operating procedure, which supply the conventions an experienced operator would otherwise carry in their head and are published with the reference implementation as a schema with worked examples.

Adoption is not currently a no-programming workflow. A prospective laboratory needs access to a programmable instrument-control path and a developer able to implement and validate the instrument driver and technique-specific data reductions (Python in our implementation). The reusable components are the checkpoint state machine, four-outcome decision interface, safety-enforcement pattern, experiment-state storage, simulated backend, and planning template. The instrument driver, safety bounds, reduction functions, and facility-specific knowledge entries must be implemented and validated locally.

This layered structure raises the question of whether the AI agent is needed at all, and whether it contributes beyond what conventional automation could do. Most decisions taken at the checkpoints of this run were determined by rule: a criterion was evaluated against the reduced data, a threshold was met or it was not, and the corresponding action followed. Twenty-one of the 35 measurements met every criterion and were released unchanged, and the three sputter-frame corrections followed from the proportional law. A conventional closed-loop implementation would reproduce these outcomes, and the replay module published with the reference implementation does exactly that from the deposited data. The agent was not substitutable in the situations that no criterion anticipated, such as attributing the emission excursion to the primary-beam path rather than to the sample, and establishing during the run the reference values that the criteria compare against.

The asymmetry between agentic control and deterministic rules is the practical argument for the approach. A deterministic workflow is straightforward to write once the events of a run are known, and the replay module is itself an example of that, written afterwards. Before the run, the set of events is not known. Anticipating and encoding every contingency demands an effort out of proportion to the study it supervises, and still leaves the session exposed to whatever was not foreseen. The flexibility of an AI agent is what allows such unforeseen situations to be resolved as they arise, and the argument is not specific to SIMS: it applies wherever a measurement consumes the specimen, so that a failure recognized only in post-run analysis cannot be repaired by repeating it.

# 6. CONCLUSIONS

In conclusion, here we developed the agentic human-AI PACE-SIMS system for an autonomous ToF SIMS and applied it to the investigation of the chemistry and stoichiometry of a series of $^{18}O$-enriched $WO_x$ films. A single ToF-SIMS session characterized four films spanning the oxide series, in both ion polarities, resolving the buried tracer layer, ranking the films by enrichment, and placing their composition on a calibrated scale: the figures and datasets presented above.

The process ran as a collaboration between the researcher and the AI agent, with authority shifting across the three phases. Planning (stating the scientific questions and quality requirements in natural language and approving the resulting plan at the two human go/no-go reviews) took roughly 20 to 30 minutes. The Run then proceeded under bounded agent autonomy for 8.1 hours, of which about forty minutes involved a human; the rest was unattended. Directed analysis and presentation, with the researcher steering the agent through reduction and figure generation, took about an hour. A working day's result came from roughly two hours of the researcher's attention.

The value of the approach becomes clearest when compared with the alternatives. Performed manually, the acquisitions alone would have required essentially a full working day of continuous researcher time at the instrument. A fixed automation script would have removed that burden, but the experiment would still have failed in two ways. First, applying a single sputter condition across the entire series, as assumed in the initial plan, would have driven three of the four films outside the required sampling range. Second, when the ion-gun emission dropped during the run, a fixed script would simply have continued repeating poor measurements and consuming additional sample locations rather than recognizing the failure and escalating it. A closed-loop implementation, in which the same acceptance criteria were coded and applied after each measurement, would have handled the first of these and much of the routine variability of the session, as discussed in Section 5.2; it is the second that separates the two approaches. In principle, some of these contingencies could be anticipated and encoded in advance. But constructing a script robust to every expected failure would require substantially more effort than specifying the experimental requirements to an agent in natural language, and it would still remain vulnerable to failures that were not anticipated beforehand, such as the emission drop encountered here.

Beyond demonstrating autonomous operation, the experiment also provided insight into the dynamic properties of the $WO_x$, in which oxygen stoichiometry controls the resistive-switching and electrochromic behavior that motivates these materials. Using the cluster-ion signal, the agent placed the four films on a calibrated composition scale, identified a 5.3% systematic offset in isotope fraction between the two detection polarities, and used the buried isotope tracer, a standard probe of oxygen transport, to infer the tracer-delivery mechanism during deposition. This interpretation was consistent with the actual growth recipe. The most metallic film also exposed the limit of the method: its cluster-ion yield was suppressed by approximately 30%, whereas the independently measured density remained physically consistent with the sputter rate. This establishes the range over which the SIMS-based composition calibration can be trusted and identifies where residual matrix effects again become important.

The demonstration also has clear limitations. The composition scale is a calibration derived from the independently measured densities and deposition information, rather than from matrix-matched standards, and is therefore validated only over the oxidized composition range examined here. The most metallic film falls outside that range. Likewise, the absolute accuracy of the isotope enrichment is limited at the few-percent level from both sides: by the inter-mode fractionation of the SIMS measurement and by the accuracy of gas delivery during deposition. Finally, the present study considers a single instrument and a single material system, and agent-based judgment inevitably introduces model dependence and potential run-to-run variability that are absent from fixed deterministic logic. The natural next steps are therefore to extend the same framework to additional feedback channels, including in-chamber metrology, to substantially longer unattended campaigns, and to other consumptive measurements where an incorrect experimental decision irreversibly spends sample material. Importantly, none of these limitations is intrinsic to the architecture of the agent-controlled approach itself.

# METHODS

## Samples

The four $WO_x$ films were grown using 100 W reactive DC magnetron sputtering (AJA Orion 8) from a 3-inch tungsten metal target in an $Ar/O_2$ mixture at a total gas flow of 40 sccm and 5 mTorr of total pressure. The oxygen flow was set to 2, 4, 6 and 8 sccm (5, 10, 15 and 20% of the total flow) for the four samples. The buried isotopic marker was produced by holding a fixed $^{18}O_2$ flow of 1 sccm during part of the deposition while the $^{16}O_2$ flow was reduced by the same amount, so that the total oxygen flow was unchanged; the argon flow was adjusted to keep the total flow at 40 sccm. The deposition rate for each film ranged from 5.9 to 7.1 nm $min^{-1}$. Films are nominally 45 nm thick and were capped in the same system with approximately 30 nm of $SiN_x$ deposited by plasma-enhanced chemical vapor deposition (Plasmatherm 790). Densities were determined independently using X-ray reflectivity (Supplementary Information, *Figure S1*).

## ToF-SIMS acquisition

Measurements were performed on a TOF.SIMS[5] - NCS instrument (IONTOF GmbH) under SurfaceLab 7.5 control, using a bismuth cluster analysis ion beam ($Bi_3^+$, 30 keV energy, 30 nA current in DC mode, ~5 μm spot size) and a cesium sputter beam (2 keV energy, 95–105 nA current, ~20 μm spot size) in non-interlaced mode, where each analysis scan with Bi liquid-metal ion gun (100 × 100 μm, 0.3 s) was followed by sputtering with Cs (350 × 350 μm, 0.5 – 4 s). A low-energy electron flood gun was used for charge compensation. Measurement positions, including locations reserved for parameter tuning and for corrective remeasurement, were fixed before the run by optical registration of the sample holder.

Each profile ran through the full stack and terminated on a dynamic stop triggered by the rise of a substrate silicon signal rather than on a fixed scan count. Because silicon is also present in the SiN cap, the trigger required a sustained rise rather than a single threshold crossing. Negative polarity (6 ns liquid-metal ion gun pulse width) tracked the oxygen isotopes together with tungsten oxide and matrix species directly. Positive polarity used cesium-cluster ($MCs_n^+$) detection, which suppresses much of the matrix dependence of secondary-ion yields, acquired in two multiplexed pulse-width phases (6 and 20 ns liquid-metal ion gun pulse width) so that the intense $Cs^+$ and $Cs_2^+$ channels and the far weaker cluster channels could each be recorded within the linear range of the detector. Pulse widths, shot counts, raster and crater sizes, tracked species with their nominal masses, stop-condition settings and the full quality-control acceptance criteria are listed in the Supplementary Information (*Section S3*).

## Reference crater-depth calibration

Sputter rates per unit ion dose were calibrated for each film in a separate operator-controlled session preceding the blind run. Craters of known ion dose were sputtered under the same beam conditions and their depths measured by atomic force microscopy (AFM) in the same vacuum chamber as the ToF-SIMS. Piecewise fit over the $SiN_x$, $WO_x$ and substrate segments yielded the depth removed per unit dose for each layer of each sample (Supplementary Information, *Figure S2*). These values, together with the composition assignment, constituted the sealed ground truth of the blind study and were not consulted until all blind-side analysis had been frozen.

## Run sequence

The 24 main measurements were acquired as two blocks with a single polarity switch between them: all negative-polarity replicates first, then all positive-polarity replicates, and finally the polarity was switched back for the closing drift-sentinel measurement. Blocking separates polarity from elapsed time, so the positive block visited the samples in reverse order, opposing sample order to time order; agreement of the sample ranking between the two blocks therefore cannot be produced by monotonic drift. Sputtering is independent of spectrometer polarity, so the per-sample sputter parameters converged during the negative block were carried unchanged into the positive block. A quality-control reference spectrum was acquired for each polarity. The reversed ordering of the positive block and the mid-block anchor measurements at spare locations on the first sample were proposed by the agent during planning rather than specified in the researcher's requirements. The session closed with a drift sentinel: a negative-polarity measurement at a spare location on the first sample, compared against the negative reference acquired at the start of the session.

## Data reduction

Depth scales were assigned per measurement by anchoring the film interfaces identified in each profile, and enrichment fractions were computed over a fixed-depth window centered on the enriched layer. All quantities reported here were regenerated in a single pass by one analysis pipeline, whose output table is archived with the dataset; earlier extractions that differ in detail are superseded by it. Details of the pipeline, the window definitions and the estimator comparisons are given in the Supplementary Information (*Section S4*).

# DATA AVAILABILITY STATEMENT

The measurement data supporting this study are deposited at https://doi.org/10.5281/zenodo.21892339. The record contains the calibrated depth profiles for all 35 measurements of the session, including the invalidated acquisitions and their remeasurements; the per-sample and per-measurement summaries; the quality-control event log; the experiment-state store and position registry; the sealed ground truth opened at unblinding, together with the pre-registered predictions and their outcomes; the AFM crater-depth and X-ray reflectivity calibration data; the full run transcript with the paired timelapse recording; and the notebook executed during the run, preserved with its outputs, whose execution cell carries the complete run log for all 35 measurements (docs/notebook/WOx_bipolar_experiment_executed.ipynb). The raw four-dimensional datasets (vendor format, of order gigabytes per measurement) are available from the corresponding author on reasonable request; every quantitative claim reported here derives from the reduced data in the deposited record.

The PACE-SIMS reference implementation is available at https://github.com/ianton86/pace-sims-reference and archived at https://doi.org/10.5281/zenodo.21906851. It contains the checkpoint-gated execution state machine, the quality-control criteria and decision framework applied at each checkpoint, the safety envelope enforced by the engine, the experiment-state schema, and a simulated instrument backend, so the full lifecycle runs without hardware; a replay module re-implements the quality-control criteria and the decision menu and, run over the logged measurement metrics of this study, reproduces the run's decision sequence.

That module was written after the run for verification and is not the code path that applied the criteria at the time, which is why the executed notebook is deposited alongside the data: the re-implementation can be compared directly against the record of what ran. Two components are not included. The instrument-communication layer wraps internal interfaces of the vendor acquisition software that are not part of its public interface and drives a shared user-facility instrument, so it is not redistributable; its release is additionally subject to ongoing technology-transfer review following the invention and software disclosures noted under Conflicts of Interest. Its public boundary is documented in the repository as an abstract driver interface, and its operation is described in the Methods and Supplementary Information (Section *2.5*). The experiment and analysis knowledge bases are also withheld, as they encode facility-specific operating procedure; the part of them on which the results here depend, the quality-control criteria, is published as executable code, and the entry schema with worked examples is provided in the repository.

# AUTHOR CONTRIBUTIONS

A.V.I. conceived the study, developed the software, performed the ToF-SIMS experiments, and analyzed the data. S.V.K. contributed to the conception, positioning, and development of the study. H.H. and Y.L. fabricated the $WO_x$ films and performed the XRR characterization. All authors wrote the manuscript.

# CONFLICTS OF INTEREST

The authors have filed invention and software disclosures with ORNL / UT-Battelle relating to the system described here. There are no other competing interests to declare.

# ACKNOWLEDGEMENTS

ToF-SIMS research and developments were supported by the Center for Nanophase Materials Sciences (CNMS), which is a US Department of Energy, Office of Science User

Facility at Oak Ridge National Laboratory. S.V.K. acknowledges support from the NSF Physics Frontier Center ATHENA (NSF 2607469).

# SUPPLEMENTARY INFORMATION

The Supplementary Information contains: the complete natural-language prompt set and checkpoint-gated run transcript, provided as a page-referenced companion file with a paired timelapse recording of the run (*Supplementary Video S1*); the blind-study predictions and the sealed ground-truth envelopes against which they were evaluated; the full acquisition parameters, tracked species and quality-control acceptance criteria; the depth-calibration pipeline, integration-window definitions and estimator comparison behind the inter-mode fractionation factor; and the composition-quantification and matrix-effect analysis recovered from the persisted raw data. Supplementary tables are numbered S1–S13.

# Supplementary Information

# PACE-SIMS: Checkpoint-Gated Autonomous SIMS Characterization with AI-Agent Quality Control

Anton V. Ievlev[1,*], Heather Hare[2], Yiyang Li[2], Sergei V. Kalinin[3]

[1] Center for Nanophase Materials Sciences, Oak Ridge National Laboratory, Oak Ridge, TN 37831

[2] Materials Science and Engineering, University of Michigan, Ann Arbor, MI 48109

[3] Materials Science and Engineering, University of Tennessee Knoxville, Knoxville, TN 37996

[*] **Corresponding author**

This document supplements the main text. Sections are numbered to match the Supplementary Information inventory. Large verbatim records — the planning-phase communication, run event log, and extraction code — are provided as companion files and indexed from the relevant section here.

Table of Content

# S1. Prompt set and gate transcripts

The autonomous run was specified entirely in natural language. No measurement script was written: the sequence, parameters, quality-control rules and safety limits were supplied as a short series of prompts, and the agent translated them into an executable notebook only after human approval at a review gate. It then executed the study over an 8.1-hour measurement session, pausing at each measurement to judge the data and to accept, correct, retry or escalate. This section documents both phases and indexes the complete turn-by-turn record, provided verbatim with timestamps as the companion file **SI_1_full_transcript.pdf**.

## S1.1. Structure of the specification

The prompt set was delivered in four parts before any plan was proposed. The opening prompt (P0) established the blind-study constraint — no memories, past conversations or stored results from prior runs on these samples were to be consulted — and directed the agent to review only the general method and to enumerate the available measurement positions without yet planning. This ordering is deliberate: it fixes the blind boundary before the agent has seen anything it could plan around.

The next three prompts supplied, in turn, the sample description and the three scientific questions (P1); the acquisition-settings baseline, given explicitly rather than drawn from stored documents (P2); and the quality-control rules, per-pause decision logic and safety limits (P3). Only then was the agent asked to propose a plan.

## S1.2. The go/no-go reviews

Two human go/no-go reviews bound the planning phase. At Gate 1 the agent returned the position registry and confirmed that no prior-run data had surfaced anywhere in the accessible tools — the registry read 144 of 144 positions fresh, with all experiment, polarity and species fields empty — establishing the blind starting condition on the record. At Gate 2 the agent proposed a complete measurement plan (species, parameters, sequence, and every deviation from the knowledge-base defaults) and did not generate the notebook until the plan was explicitly approved.

The Gate-2 exchange was not a formality. Over twelve turns the proposed plan was corrected in ways that materially affected validity: a stage z-range that would have locked out every registered position was fixed; crater sizes and imaging resolution were made global after the agent argued that a converged sputter-frame count transfers between tuning and acquisition only if the sputter current density is identical; the replicate ordering was changed to position-major so that a per-sample rate correction could not land between replicates of the same sample; the negative measurement was simplified to a non-multiplexed acquisition; and the sample-to-position mapping was confirmed explicitly, since nothing in the data would reveal a mis-assignment. This back-and-forth is the clearest evidence in the study for the first thesis of the main text — that specifying a measurement in natural language collapses the effort of encoding it and surfaces design errors before any beam time is spent — and it is reproduced in full in the companion file.

## S1.3. The complete transcript

The full run is provided verbatim as the companion file **SI_1_full_transcript.pdf** (37 pp.): every human prompt and agent response, in order, with instrument-clock and elapsed (T+) timestamps, from the opening blind-study constraint through the end-of-run compilation. The post-run directed-analysis session is documented separately (*Sections S4, S5*) and is not included there. Obvious spelling errors in the human prompts have been silently corrected; content is otherwise unaltered. Email addresses and absolute file paths are redacted. Colored flags in the left margin mark pipeline milestones (tuning convergence, polarity switch, block completion, shutdown), the single LMIG-emission fault, the three per-sample recalibrations, and each mid-run human intervention. The record divides into three phases, located in the companion file as follows (*Table S1*).

A timelapse recording of the same run is provided as Video S1. It condenses the full measurement session — stage motion between sample positions, the optical view at each location, and the live depth-profile and mass-spectrum panels — into a short visual companion to the textual transcript below, showing the same sequence of measurements, recalibrations and the polarity switch as they unfolded at the instrument. The video's elapsed-time counter (T+0:00 at notebook execution) is reproduced as the T+ column in the transcript and its event index below, so any moment in the video maps directly to the corresponding turn.

***Table S1.*** *Phase index to the companion transcript file, with page ranges.*

| Pages | Phase | Content |
|---|---|---|
| pp. 1–14 | Planning | P0–P3 prompt set, Gate 1 registry check, twelve-turn Gate 2 plan review, approval, notebook generation |
| pp. 14–34 | Run | Checkpoint-gated execution: 35 |
| | | measurements, each followed by a QC decision at its pause; three unscripted recalibrations; LMIG-dropout diagnosis; polarity switch; drift sentinel; shutdown |
| pp. 35–37 | Report | End-of-run compilation of enrichment fractions across all samples and both polarities |

The measurement phase ran 8.1 h on the instrument clock (09:06–17:13), from the first tuning acquisition to instrument shutdown; 35 measurements in total. The wider session — planning, the checkpoint-gated run, and the end-of-run compilation — spans the timestamps in the companion file. The three recalibrations and the LMIG-fault diagnosis appear in the run phase as pauses where the agent departed from routine acceptance — the events on which the main text's second thesis, that runtime judgment outperforms fixed rules, is evidenced.

## S1.4. Locating the main events

The table below gives the page in the companion file at which each flagged event appears, so a reader can turn directly to any milestone, fault, recalibration or human intervention and read the surrounding exchange (*Table S2*). T+ is elapsed time from run start (T+0:00 = notebook execution, matching the Video S1 counter); times are on the instrument (local) clock, as shown in the transcript headers. The same events carry colored margin flags in the file.

***Table S2.*** *Locator for each flagged run event: page, elapsed time, type and description.*

| Page | T+ | Time (local) | Type | Event |
|---|---|---|---|---|
| p. 17 | T+0:29 | 09:34 | Tuning | Tuning converged — negative QC reference set |
| p. 18 | T+0:59 | 10:04 | Fault | LMIG emission dropout (10:04 instrument clock) |
| p. 19 | T+1:02 | 10:07 | Human | QC routine extended to xz/yz interface maps |
| p. 21 | T+1:12 | 10:17 | Human | LMIG realignment note |
| p. 21 | T+1:12 | 10:17 | Recalibration | S1 rep3 redo — post-dropout source recovery |
| p. 22 | T+1:20 | 10:25 | Human | Reverse positive block (S4→S1) to decorrelate Cs drift |
| p. 23 | T+1:34 | 10:39 | Recalibration | S2 recalibration — 11→10 frames |
| p. 24 | T+1:47 | 10:52 | Human | Crater-alignment QC gap flagged |
| p. 26 | T+2:29 | 11:34 | Recalibration | S3 recalibration — 11→13 frames |
| p. 27 | T+3:14 | 12:19 | Recalibration | S4 recalibration — 11→9 frames |
| p. 28 | T+3:47 | 12:52 | Milestone | Negative block complete (12 main measurements) |
| p. 28 | T+4:07 | 13:12 | Switch | Polarity switch NEG→POS |
| p. 32 | T+6:40 | 15:45 | Human | Full-autonomy handoff — run to completion |
| p. 34 | T+8:08 | 17:13 | Milestone | Drift sentinel measured → shutdown released |
| p. 35 | T+8:17 | 17:22 | Milestone | End-of-run compilation of enrichment fractions |

The three recalibrations (S2, S3, S4) and the single LMIG fault are the events on which the main text’s account of runtime judgment rests; each is a point where the agent departed from routine acceptance. The S1 rep3 redo is a recovery from the fault rather than a per-sample rate correction, and is listed separately.

Companion file: **SI_1_full_transcript.pdf** (37 pp., timestamped, T+ elapsed clock, event-flagged).

# S2. Blind validation: predictions and sealed ground truth

Before the autonomous run began, four quantitative predictions — each with an acceptance threshold set in advance at the few-percent level — were stated, and the samples were assigned to holder positions in a randomized, operator-confirmed order that carried no information the agent could exploit (*Table S3*). The prompt set that specified the run is reproduced in *Section S1*; the predictions themselves are stated below.

## S2.1. The four predictions

Each prediction was chosen to test a different aspect of the agent's in-run behavior against an external check rather than against its own internal consistency.

***Table S3.*** *The four pre-registered predictions and their pass/fail check.*

| # | Prediction | Check |
|---|---|---|
| i | The relative sputter rates inferred by the agent during the run would agree with the sealed AFM calibration. | Independent metrology (Envelope M) |
| ii | The mass removed per unit ion dose would be constant across the four films, at the level of constancy already observed on the calibration dataset. | Mass conservation |
| iii | No recalibration would be required in the second polarity block, because the corrections established in the first block transfer to it. | Physical assumption behind the corrections |
| iv | A drift sentinel acquired at the end of the session would reproduce the measurand within the run's own quality-control tolerance. | Temporal-artifact bound |

The acceptance threshold for each was set in advance at the few-percent level. The four are deliberately different in kind: predictions (i) and (ii) ask whether the agent's adaptive corrections are physically correct rather than merely self-consistent, testing them against independent metrology and against mass conservation; (iii) tests the physical assumption underlying those corrections; and (iv) bounds how much of any difference between samples could be an artifact of when it was measured.

## S2.2. The sealed ground-truth envelopes

Two documents recorded the ground truth against which the blind run was validated. Both existed before the autonomous run and were withheld from the agent throughout planning, execution and the directed analysis; they were opened only at unblinding. Envelope C fixes the identity and composition of each sample (*Table S4, Figure S1*); Envelope M provides an independent AFM crater-depth calibration of the film and cap thicknesses and sputter rates (*Table S5, Figure S2*). Because the mounting order was randomized and operator-confirmed at mounting, the sample-to-position mapping is itself part of the sealed record: nothing in the SIMS data could recover it. Both envelopes are reproduced verbatim below.

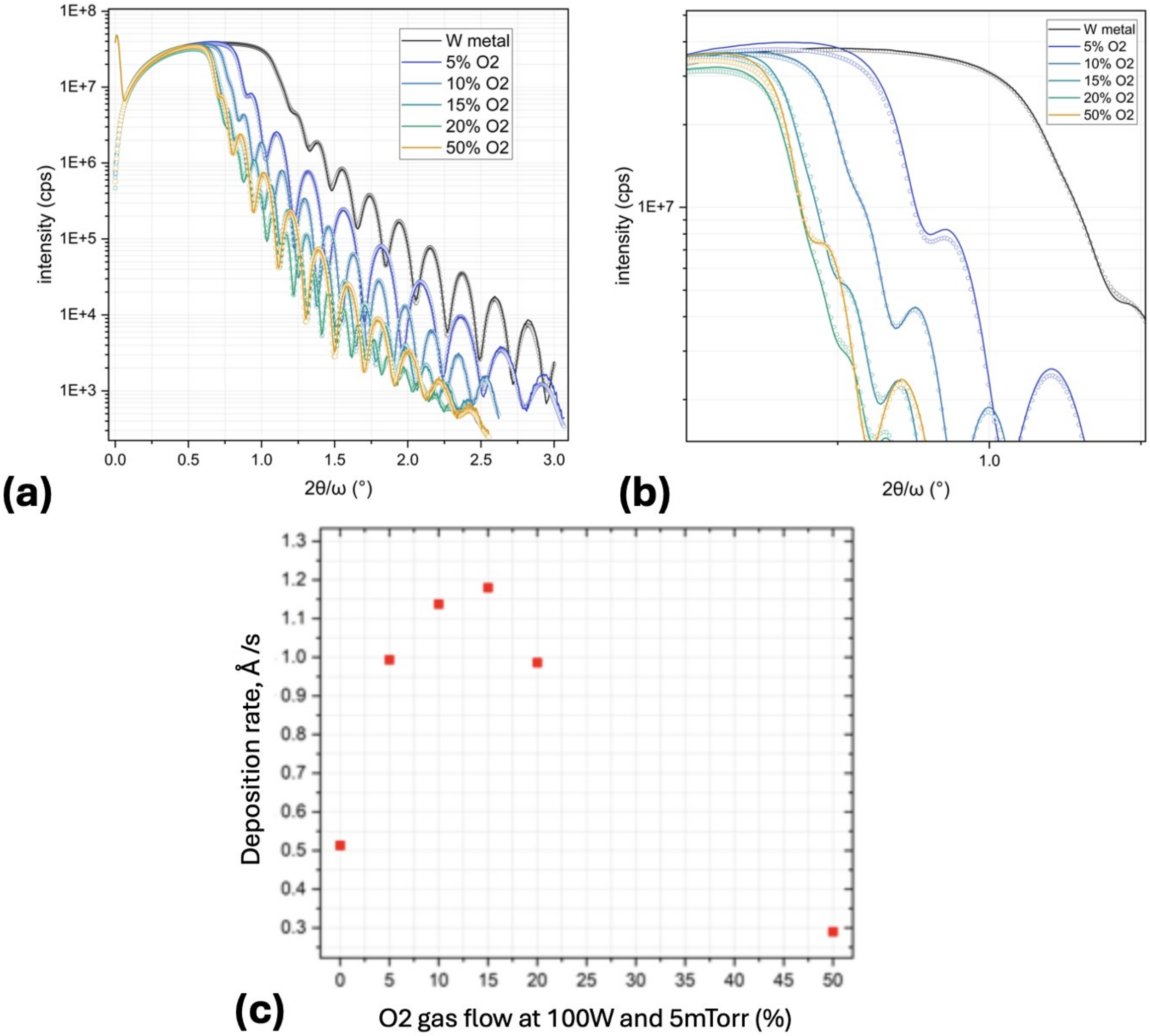

***Figure S1.*** *Results of X-ray reflectometry measurements on the grown $WO_x$ films. (a) Full scale and (b) zoom in on critical angles; (c) Fitted data of deposition rates. Higher critical angles imply greater difference in the electron density ($n_e$ $nm^{-3}$)*

*between the $WO_x$ film and the Si substrate; this electron density is used to compute the density. The periodicity relates to the thickness. Intensity is given in counts per second.*

***Table S4.*** *Envelope C: sealed sample identities, densities, deposition rates and derived compositions.*

| Registry label | $O_2$ flow during deposition (sccm) | Density (g/cc) | Deposition rate (nm/min) | Mixing-model x in $WO_x$ |
|---|---|---|---|---|
| S1 (POS1) | 4 (10%) | 8.16 | 6.83 | ≈ 1.9 |
| S2 (POS2) | 6 (15%) | 6.85 | 7.08 | ≈ 2.7 |
| S3 (POS3) | 2 (5%) | 10.3 | 5.96 | ≈ 1.1 |
| S4 (POS4) | 8 (20%) | 6.51 | 5.92 | ≈ 2.96 |

Deposition-series endpoints for reference: 0 % → 18.7 g/cc (metallic W); 50 % → 6.46 g/cc (saturated, ≈ amorphous $WO_3$). Mounting order was randomized (labels do not follow the fabrication series); POS↔sample mapping operator-confirmed at mounting.

Sealed prior to mounting; opened at unblinding. The mixing-model x values are derived by the depositor from the deposition flow and density series, not from any SIMS measurement — a point the main text is careful to preserve, since the SIMS composition scale is calibrated against these values rather than establishing them independently.

Dataset 20260723_WOx_thickness_calibration, sealed 2026-07, opened at unblinding. Method: per-sample ToF-SIMS crater series with AFM depths, piecewise global fit ($SiN_x$ / $WO_x$ / Si); a representative calibration is shown in *Figure S2*. Fabrication labels (WOx5–WOx20) are the deposition flow percentages; the mapping to registry labels is given in Envelope C.

***Table S5.*** *Envelope M: sealed AFM crater-depth calibration per sample.*

| Sample | $SiN_x$ (nm) | $WO_x$ (nm) | SiN nm/$10^{15}$ | $WO_x$ nm/$10^{15}$ | $WO_x$ nm/scan |
|---|---|---|---|---|---|
| WOx5 | 29.0 ± 0.5 | 44.2 ± 3.5 | 0.87 | 0.92 | 1.18 |
| WOx10 | 28.8 ± 1.0 | 45.9 ± 3.0 | 0.87 | 1.12 | 1.42 |
| WOx15 | 29.7 ± 0.5 | 45.3 ± 3.0 | 0.87 | 1.26 | 1.63 |
| WOx20 | 30.0 ± 0.8 | 46.5 ± 3.0 | 0.88 | 1.41 | 1.81 |

Key sealed facts: $WO_x$ thickness = 45 nm within error on all samples (no resolved trend); SiN cap common (29–30 nm) with rate constant to 0.87–0.88 nm/$10^{15}$ across all four samples (a four-fold method validation); $WO_x$ per-dose rate monotonic with $O_2$ flow; mass-removal closure, rate × ρ, constant to ±4.5 % on this dataset.

These two records are the sealed truth referenced throughout *Section* of the main text: Envelope C supplies the flow order and composition that three independent SIMS observables reproduce, and Envelope M supplies the thickness and rate calibration against which the agent's in-run sputter-rate inferences agree to about one percent.

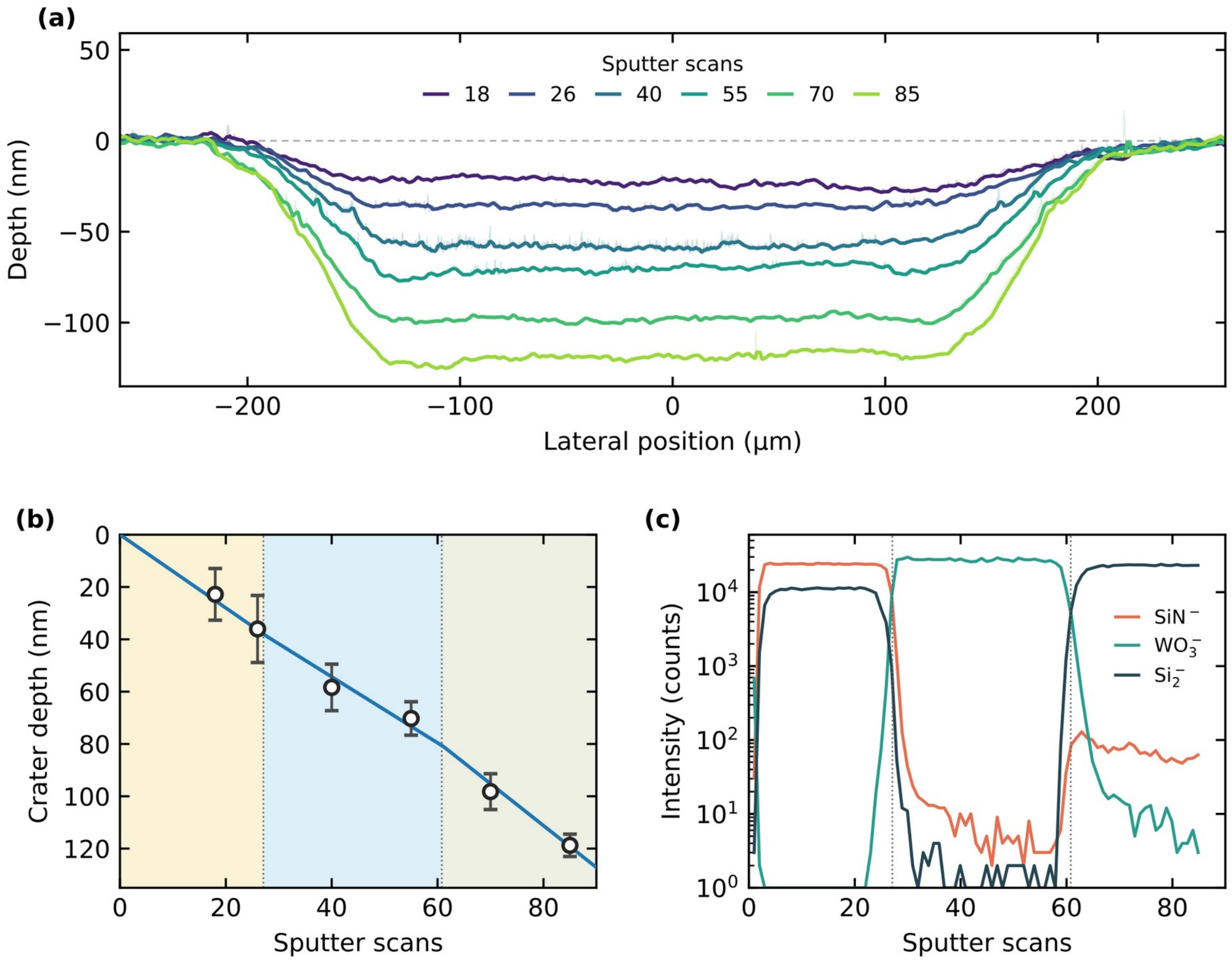

***Figure S2.*** *Representative AFM crater-depth calibration, shown for sample $WO_x$ 10% $O_2$ flow; the other samples were calibrated by the same procedure. (a) AFM line profiles across the sputter crater after increasing numbers of sputter scans (18–85), giving the crater depth at each dose. (b) Crater depth versus sputter scans with the piecewise-linear fit used for calibration; the shaded bands mark the SiN cap, $WO_x$ film and Si substrate regimes, whose boundaries are set by the co-registered depth profile in (c). (c) The corresponding ToF-SIMS depth profile ($SiN^-$, $WO_3^-$, $Si_2^-$), which locates the cap/film and film/substrate interfaces that anchor the fit boundaries in (b). Depths are AFM-measured; per-sample thicknesses and sputter rates are tabulated in Table S5.*

## S2.3. Outcome at unblinding

At unblinding, each prediction was compared against the sealed record and the run data. All four were confirmed inside their registered thresholds; the outcomes are summarized in *Table S6* and reported in full in *Section 5* of the main text.

***Table S6.*** *Predictions evaluated at unblinding: registered criterion, outcome and result.*

| # | Registered acceptance criterion | Outcome at unblinding | Result |
|---|---|---|---|
| i | Agent-inferred relative sputter rates agree with sealed AFM calibration; registered band ±3%. | Single conversion factor fits all four samples with a 1.05% spread (Figure 7b). | Confirmed |
| ii | Mass removed per unit ion dose constant across the four films, at the constancy of the calibration set; registered band ±5%. | Volumetric rate × film density constant to ±3.1% (SD) on blind data (Figure 7c). | Confirmed |
| iii | No recalibration required in the second polarity block; corrections from the first block transfer. | Positive block ran on the carried-forward per-sample frames; zero recalibrations, every band check passed. | Confirmed |
| iv | End-of-session drift sentinel reproduces the measurand within the run's own QC tolerance. | $^{18}O$ fraction 7.93% vs 7.74% reference — +2.5% relative, 1.8σ within the S1 spread, while absolute yield moved 12%. | Confirmed |

# S3. Acquisition parameters and quality-control criteria

This section lists the acquisition configuration used for every measurement in the run and the quantitative acceptance criteria applied at each checkpoint. The values were fixed in the approved plan (*Section S1*) and, except for the per-sample sputter-frame count, were identical across all measurements. They are collected here so that the run is fully specified without recourse to the transcript.

## S3.1. Common acquisition settings

Measurement mode 3D; imaging resolution 64×64; primary (analysis) crater 100 µm; sputter crater 350 µm; non-interlaced sputtering; peak margin 0.25 u; acceptance-circle adjustment enabled; sputter-crater adjustment disabled; electron flood gun on. The primary-ion species ($Bi_3^+$), buncher state and instrument cycle time were inherited from the manual instrument state; the cycle time was confirmed at run time to pass a mass range of 517 u, placing $WCs_2^+$ (449.76 u) in range.

## S3.2. Tracked species

The negative polarity used a single non-multiplexed acquisition (LMIG pulse width 6 ns, 4 shots per pixel). The positive polarity used a two-phase multiplexed acquisition in each measurement: phase A (chopper 6 ns, 4 shots per pixel), in which $Cs^+$ is quantitative, and phase B (chopper 20 ns, 8 shots per pixel). $Cs_2^+$ was recorded in both phases to link them. The tracked negative-polarity peaks and their nominal masses are listed in *Table S7*, and the positive-polarity peaks in *Table S8:*

**Negative polarity (6 ns, 4 shots/pixel):**

***Table S7.*** *Tracked negative-polarity species with nominal masses and roles.*

| Species | Nominal mass (u) | Role |
|---|---|---|
| $^{16}O^-$ | 15.9949 | major oxygen |
| $OH^-$ | 17.0027 | interference diagnostic at m/z 18 |
| $^{18}O^-$ | 17.9992 | tracer |
| $Si^-$ | 27.9769 | stop trigger; substrate anchor |
| $^{16}O_2^-$ | 31.9898 | m/z-32 denominator (saturation-immune route) |
| $SiN^-$ | 41.9800 | cap normaliser; cap/substrate discrimination |
| $WO_3^-$ | 231.9356 | tungsten matrix |

**Positive polarity (phase A 6 ns/4; phase B 20 ns/8):**

***Table S8.*** *Tracked positive-polarity species with nominal masses, phase and roles.*

| Species | Nominal mass (u) | Phase | Role |
|---|---|---|---|
| $Cs^+$ | 132.9054 | A | quantitative in phase A only |
| $Cs_2^+$ | 265.8109 | A, B | cross-phase link; positive normaliser |
| $SiCs^+$ | 160.8824 | B | stop trigger; substrate anchor |
| $^{16}OCs^+$ | 148.9004 | B | O route (cross-check) |
| $^{18}OCs^+$ | 150.9046 | B | O route (found interfered; excluded) |
| $^{16}OCs_2^+$ | 281.8058 | B | primary O quantification |
| $^{18}OCs_2^+$ | 283.8101 | B | primary O quantification (tracer) |
| $WCs^+$ | 316.8564 | B | tungsten composition probe |
| $WCs_2^+$ | 449.7618 | B | tungsten composition probe |

The m/z-34 channel was excluded from live tracking by instruction and extracted from the raw data during analysis. $^{18}OCs^+$ was retained as a deliberate redundancy; it proved to be interfered by the Cs–water adduct and was excluded from quantification, having flagged itself rather than biasing a result.

## S3.3. Dynamic stop condition

Both polarities used a dynamic depth-profiling stop on a silicon-family peak — $Si^-$ in negative, $SiCs^+$ (phase B) in positive. Because silicon is present from the surface in the SiN cap, the trigger was set to fire on the second threshold crossing (the substrate rise) rather than the first (the cap): threshold 400 counts, trigger count 2, no ignored initial scans, 30 post-trigger scans, and a 400-scan backstop that was never reached. The threshold was fixed at run time from the measured cap and substrate silicon levels during the reconnaissance tuning acquisition.

## S3.4. Per-sample sputter-frame counts

The sputter-frame count was the one parameter that varied between samples, tuned so that each profile fell within the 50–60 point acceptance band across the $WO_x$ layer. Starting from a 45-frame reconnaissance on S1, the count was converged by the proportional rule new = current × measured/55 and then carried unchanged from the negative block into the positive block. The four samples required distinctly different values — a 44% spread — so a single tuned value would have placed three of the four outside the band (*Table S9*).

***Table S9.*** *Converged per-sample sputter-frame counts.*

| Sample | Converged sputter frames |
|---|---|
| S1 | 11 |
| S2 | 10 |
| S3 | 13 |
| S4 | 9 |

## S3.5. Quality-control acceptance criteria

At each checkpoint the agent evaluated the completed measurement against the criteria below, applied against the same-polarity reference. Composition differences between samples were expected and were never treated as failures (*Table S10*).

***Table S10.*** *Quality-control acceptance criteria and the action taken on each failure.*

| Check | Acceptance | Action on failure |
|---|---|---|
| Sampling density | 50–60 points across the $WO_x$ layer | correct sputter frames (proportional rule), invalidate, redo at a spare, carry the corrected value forward |
| Saturation | per-shot rate <5 counts/pixel/shot on any channel used for ratios (ceilings 81 920 counts/scan negative and phase A; 163 840 phase B) | hold and escalate (no in-plan remedy: pulse widths were fixed) |
| Stop termination | terminate on the dynamic trigger, not the backstop; ≥30 scans of substrate plateau; no trigger inside the cap | correct the stop parameters, invalidate, redo at a spare |
| Profile shape | flat interfaces, no mid-film discontinuity, no crater-edge or tilt artifact (xz/yz cross-sections; lateral uniformity $U = \text{std}/\sqrt{\text{mean}}$, threshold 1.3) | one redo unchanged, then flag and proceed |
| Absolute yield | uniform scaling with ratios, thickness and geometry intact | flag and proceed (ratios are insensitive to it) |
| Channel integrity | every peak present and non-empty | hold and escalate |

A second failure on the same measurement, or any condition outside this table, escalated to the operator and held rather than being auto-corrected. The distinction between a uniform yield change (passed) and a change in profile shape (one redo, then flagged) reflects that the study's deliverables are isotope ratios and layer depths, which a uniform scaling leaves intact.

# S4. Depth-calibration pipeline and window analysis

This section documents the post-run data-reduction pipeline: how depth profiles were calibrated to a common depth scale, how the enriched-layer integration windows were defined, and the estimator comparison that underlies the inter-mode isotope-fractionation factor reported in the main text. All results in this section were obtained by re-interrogating the persisted raw data from the run, with no additional acquisition.

## S4.1. Depth calibration

Each depth profile was calibrated on its own interfaces using a two-layer model: a SiN capping layer (nominal 30 nm) above the $WO_x$ film (nominal 45 nm). The cap/film and film/substrate boundaries were located per measurement from the $WO_3^-$ rise and fall in negative polarity and the $SiCs^+$ fall and rise in positive polarity, and the scan axis was scaled to depth against these two known thicknesses. Because the sputter-frame count differed between samples, this per-measurement anchoring is what makes the four depth scales comparable.

The capping layer provides an internal check. SiN is nominally the same material on all four samples, so its sputter rate per frame should be constant regardless of the underlying film. Measured across the four samples, four different frame settings, and hours of run time, it is (*Table S11*):

***Table S11.*** *SiN-cap sputter-rate constancy validating the depth calibration.*

| Sample | Frames | Cap scans | SiN nm/frame | $WO_x$ nm/frame |
|---|---|---|---|---|
| S1 | 11 | 47 | 0.0580 | 0.0731 |
| S2 | 10 | 52 | 0.0577 | 0.0865 |
| S3 | 13 | 39 | 0.0592 | 0.0618 |
| S4 | 9 | 57 | 0.0585 | 0.0926 |

**SiN erosion rate: 0.0584 ± 0.0006 nm/frame — a 1.1% spread.** This constancy independently validates the depth calibration, the 30 nm cap thickness, and the linearity of the frame-to-rate scaling that underlies the proportional tuning rule used during the run. The $WO_x$ rate, by contrast, varies 1.5× across the series and is monotonic with enrichment, which is a sample property (*Section S5*), not a calibration artifact.

## S4.2. Enriched-layer integration windows

The reported isotope fractions are integrated over the enriched layer, not the whole film, so a window must be defined. Three windows were examined to characterize the sensitivity of the result to that choice: a 10 nm fixed window on the flat top of the tracer band (47.5 – 57.5 nm from the surface), the same 10 nm width re-centred on each profile's own band, and a wider 16 nm window (44.5 – 60.5 nm) that admits more of the unenriched film on either side.

The fixed and layer-centred windows agree on the negative profiles — whose band centre already coincides with the fixed window — but diverge on the positive profiles, because the positive band centre sits about 1.8 nm shallower than the negative one. A window fixed in absolute depth is therefore slightly mis-centred on the positive profile and clips one flank. This registration offset, not any difference in band shape, is what makes the apparent inter-mode ratio depend on the window: the peaking factor (10 nm ÷ 16 nm integrated fraction) is 1.115 negative and 1.120 positive — the two polarities see essentially the same band shape to within 0.5%.

Widening the window lowers the reported fraction on every sample, since it dilutes the enriched core with unenriched film; the effect is uniform across samples and polarities and does not change the ranking. The window is thus a reproducible convention rather than a source of systematic error, provided the same window is applied to both polarities and the registration offset is accounted for.

## S4.3. Estimator comparison and inter-mode fractionation factor

The main text reports a single inter-mode fractionation factor — the systematic ratio between the positive (cluster-ion) and negative (atomic-ion) isotope fractions — that is constant across the enrichment range. Arriving at that figure requires choosing an estimator that is not biased by the depth-registration offset of *Section S4.2*. Four estimators were compared across all four samples (*Table S12*):

***Table S12.*** *Inter-mode fraction ratio for four estimators across the four samples.*

| Estimator | S3 | S1 | S2 | S4 | Mean pos/neg |
|---|---|---|---|---|---|
| 10 nm fixed window | 1.0235 | 1.0264 | 1.0414 | 1.0189 | +2.8 ± 1.0% |
| 16 nm fixed window | 1.0148 | 1.0380 | 1.0353 | 1.0027 | +2.3 ± 1.7% |
| 10 nm layer-centred | 1.0469 | 1.0515 | 1.0761 | 1.0583 | +5.8 ± 1.3% |
| Peak fraction | 1.0604 | 1.0456 | 1.0537 | 1.0288 | +5.3 ± 0.6% |

The fixed-window estimators read low (+2–3%) precisely because they are mis-centred on the positive profile and clip its flank; the layer-centred and peak-fraction estimators, which follow each profile's own band, read +5–6%. Among these the peak-fraction estimator is the tightest and the least sensitive to window placement, and it gives a factor that is constant across the full enrichment range sampled here ($f = 0.12$–$0.47$):

**Inter-mode fractionation factor: +5.3 ± 0.6%.** This is the figure quoted in the main text. The window-based numbers above are reported only to show how the estimator choice, and specifically the registration offset, moves the apparent value; they are illustration, not competing measurements. An earlier, blind-time extraction had read the offset as growing when the window narrowed, and inferred from that a difference in band shape between the two modes and hence "no single bias factor"; the re-extraction on the frozen dataset removes the registration artifact and recovers a single, constant factor.

# S5. Composition quantification and matrix effects

This section provides the supporting analysis for the absolute-composition results in the main text: the derivation of the composition parameter from the independently measured densities, the calibration of the O/W cluster-ion ratio against it, the metallic-film outlier that bounds the calibration, and the secondary matrix-effect and tracer-delivery findings recovered from the same dataset.

## S5.1. Composition from measured density

Absolute composition cannot be read directly from the secondary-ion signal, because secondary-ion yields depend on the matrix that emits them and no matrix-matched standards exist for a graded oxide series. The independently measured film densities take the place of standards. Each film is treated as a volume-additive mixture of metallic tungsten and stoichiometric $WO_3$, with endpoint densities $\rho_W = 18.70$ and $\rho_{ox} = 6.46$ g/cm$^3$. The oxide mass fraction follows from the measured density ρ as

$$w = (1/\rho - 1/\rho_W) / (1/\rho_{ox} - 1/\rho_W) \quad \text{(S1)}$$

Mass fraction is then converted to a composition parameter by atom bookkeeping, not by scaling the mass fraction directly. Using molar weights $M(WO_3) = 231.84$ and $M(W) = 183.84$ g/mol, the molar amounts are $n_{ox} \propto w/M(WO_3)$ and $n_W \propto (1 - w)/M(W)$, and the composition is

$$x = 3 \cdot n_{ox} / (n_{ox} + n_W) \quad \text{(S2)}$$

Here the factor of three counts the three oxygen atoms per oxide formula unit. The atom-bookkeeping step matters: x is not three times w, because the two molar masses differ by 26%, and the shortcut $x = 3w$ overstates the composition of the metallic film by roughly 15%.

The resulting values, derived here from the measured densities rather than taken from the sealed record, are given in *Table S13*:

***Table S13.*** *Composition parameter x derived from each measured film density, with the measured Cs-cluster ratio against which it is calibrated.*

| Sample | ρ (g/cm$^3$) | Oxide vol. frac. φ | Mass frac. w | x (atom bookkeeping) | $OCs_2^+/WCs^+$ (measured) |
|---|---|---|---|---|---|
| S3 | 10.30 | 0.686 | 0.430 | 1.124 | 6.33 ± 0.05 |
| S1 | 8.16 | 0.861 | 0.682 | 1.888 | 15.02 ± 0.11 |
| S2 | 6.85 | 0.968 | 0.913 | 2.678 | 22.05 ± 0.28 |

| S4 | 6.51 | 0.996 | 0.988 | 2.956 | 23.97 ± 0.53 |
|---|---|---|---|---|---|

These x values are the independent composition scale against which the ion-yield ratio is calibrated in *Section S5.2*. They are consistent with the sealed depositor values (Envelope C: 1.10, 1.90, 2.70, 2.96) to within the density measurement, but are derived here from the densities so that the calibration rests on a measured quantity rather than on the sealed record.

## S5.2. O/W calibration and the metallic-film boundary

Of the two detection modes only the positive one is a candidate for absolute composition: the tungsten oxide anion $WO_3^-$ does not track composition monotonically, whereas the Cs-cluster tungsten signal does. The composition probe is therefore the cluster-ion ratio $OCs_2^+/WCs^+$, calibrated against the density-derived x of *Section S5.1*. Zero oxygen signal is expected at zero oxide, so the physical origin serves as an additional anchor.

**A line through the origin and the three oxidized films (S1, S2, S4) is O/W = 8.16 x, with $R^2$ = 0.9995.** The most metallic film, S3, falls about 30% below this line and is excluded from the fit. This is a calibration — the ratio is fitted to an independently established composition scale — not a standard-free determination of x: the ratio has no absolute scale of its own, since it depends on useful yields, transmission and interval widths, and acquires one only through the density-derived composition. The through-origin form makes the calibration usable to convert O/W to x for further samples on this stack at these settings, within the oxidized range it spans.

S3 marks the boundary of that range. Its 30% shortfall is not a composition error: the mass-removal closure (rate × density) holds for S3 at its measured density, so the deficit is in the ion yield, not the amount of material removed. At x ≈ 1.1 the film is near-metallic, and the residual matrix effect that the Cs-cluster mode suppresses but does not eliminate re-emerges. The calibration is therefore trustworthy across the oxidized range (x ≈ 1.9–3.0) and should not be extrapolated to near-metallic compositions.

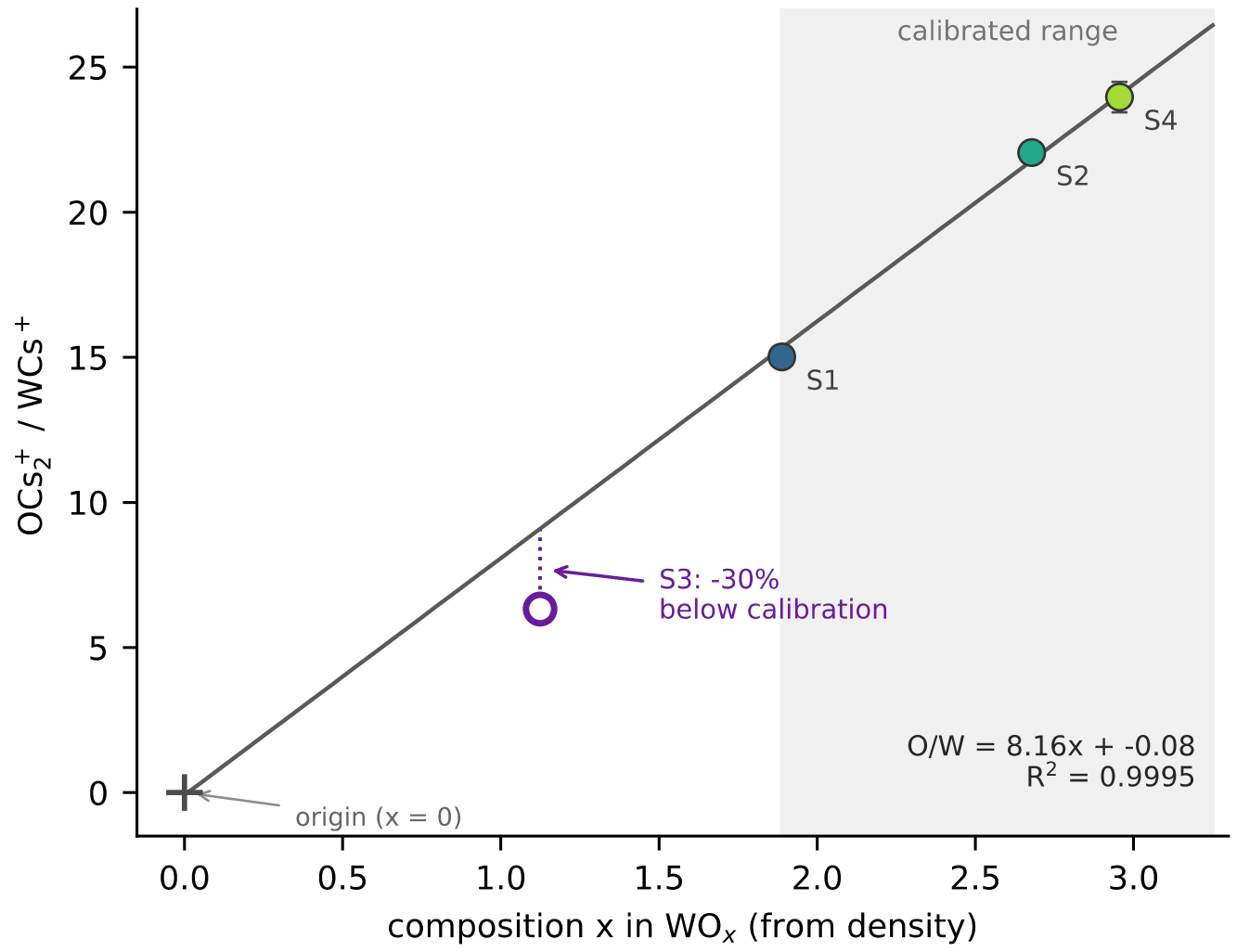

***Figure S3.*** *Composition calibration from the O/W cluster ratio. The composition x of each film is derived from its measured density (Table S13) and the working ratio $OCs_2^+/WCs^+$ is regressed against it. The line is fitted to the physical origin (zero oxygen, zero signal) and the three oxidized films; the most metallic film, S3, falls about 30% below the trend and is excluded from the fit, marking the boundary of the calibrated range. Error bars are 1 SD over three replicates. This figure was presented as Figure 8 of the submitted manuscript and is retained here.*

## S5.3. Residual matrix effects

Two secondary observations show the matrix effect operating where it is not suppressed. First, the raw tungsten cluster signal $WCs^+$ (normalized to $Cs^+$) spans 3.3× across the series and reproduces the enrichment ranking exactly, while the tungsten oxide anion $WO_3^-$ (normalized to $SiN^-$) spans only 1.8× and orders the samples differently, anti-correlated with oxygen content. The anion tracks local oxidation state rather than tungsten content; the Cs-cluster cation suppresses that dependence and tracks tungsten more faithfully. The 3.3× span of $WCs^+$ against the 1.8× span of the true tungsten density is itself a residual matrix effect — the cluster mode is more faithful than the anion, not perfectly matrix-independent.

Second, the $Cs_2^+$ dimer signal varies 2.1× between samples (from about 680 counts on S3 to 290 on S4) while the $Cs^+$ monomer stays flat at about 46 000. $Cs_2^+$ formation requires two cesium atoms to associate and so depends on surface cesium coverage, which varies with the matrix; $Cs^+$ is re-sputtered implanted cesium and tracks the beam. The dimer is therefore itself a composition-sensitive probe rather than a normalizer, and was not used as one.

## S5.4. Positive-polarity outlier

The gas-phase isotopic composition set by the deposition recipe (50.0, 25.0, 16.7 and 12.5% for the four samples) transfers to the film, because the two oxygen isotopes share the same reaction pathways. Measured peak fractions read 94 to 99% of the gas values in negative polarity and 100 to 108% in positive, bounding both modes at the few-percent level. Within this, one positive-polarity point sits above its gas value by more than the others. The provisional attribution is mass-flow-controller tolerance in the deposition: the recipe fixes the intended flow, but the realized flow carries the controller's tolerance, and a small flow error moves the delivered isotope fraction (*Section S5.5*) without any measurement error. The attribution is provisional because the run has no independent record of the realized flows; it is the most parsimonious explanation given that the negative mode, measured on the same films, does not show the same excursion.

## S5.5. Tracer-delivery mechanism

The buried tracer was delivered by displacement, not addition. Across the four samples the product of the measured isotope fraction and the oxygen flow is constant — $f_{18} \times (O_2$ flow$) = 232 \pm 6$, constant to 2.5% in negative polarity and $238 \pm 8$ (3.4%) independently in positive — whereas the product of the isotope fraction and the composition x scatters by 16%. Equivalently, the isotope fraction follows an inverse flow law, $f_{18} \propto$ flow^(−0.99 ± 0.03), indistinguishable from exactly 1/flow.

This constrains the deposition process even though SIMS measures only an isotope ratio. The two oxygen isotopes have essentially identical sticking, dissociation and reaction probabilities, so the gas-phase isotope ratio transfers to the film unchanged regardless of what fraction of arriving oxygen is incorporated; stoichiometry has no such invariance, since x depends on sticking, ion energy, resputtering and temperature. That asymmetry is why $f_{18}$ × flow holds to 2.5% while $f_{18} \times x$ scatters by 16%: the isotope fraction is a near-direct readout of a gas-phase setpoint. The inverse-flow dependence indicates a fixed $^{18}O_2$ supply displacing a varying $^{16}O_2$ supply, and the inference was subsequently confirmed against the growth recipe. The relative scatter quoted here is how well the 1/flow law holds; it is not a fraction of $^{18}O$ consumed, which is not recoverable from a ratio measurement.